\documentclass[preprint2]{aastex61}

\usepackage{natbib}
\usepackage{amsmath}
\received{April 25, 2018}
\revised{May 2, 2018}
\accepted{May 3, 2018}
\submitjournal{AJ}

\shorttitle{2014 MU69 Orbit Fitting}
\shortauthors{Porter et al.}

\begin{document}

\title{High-Precision Orbit Fitting and Uncertainty Analysis of (486958) 2014 MU69}

\correspondingauthor{Simon B. Porter}
\email{porter@boulder.swri.edu}

\author[0000-0003-0333-6055]{Simon B. Porter}
\author[0000-0003-0854-745X]{Marc W. Buie}
\author[0000-0002-6722-0994]{Alex H. Parker}
\author{John R. Spencer}
\affil{Southwest Research Institute, 1050 Walnut St. Suite 300, Boulder, CO 80302, USA}

\author[0000-0001-8821-5927]{Susan Benecchi}
\affil{Planetary Science Institute, 1700 East Fort Lowell, Suite 106, Tucson, AZ 85719, USA}

\author[0000-0002-2718-997X]{Paolo Tanga}
\affil{Universit\'{e} C\^{o}te d'Azur, Observatoire de la C\^{o}te d'Azur, CNRS, Laboratoire Lagrange,
Bd de l'Observatoire, CS 34229, 06304 Nice cedex 4, France}

\author[0000-0002-3323-9304]{Anne Verbiscer}
\affil{University of Virginia, P.O. Box 400325, Charlottesville, VA 22904, USA}

\author[0000-0001-7032-5255]{J. J. Kavelaars}
\author{Stephen D. J. Gwyn}
\affil{National Research Council of Canada, Victoria, BC, Canada}

\author[0000-0001-8242-1076]{Eliot F.  Young}
\affil{Southwest Research Institute, 1050 Walnut St. Suite 300, Boulder, CO 80302, USA}

\author[0000-0003-0951-7762]{H. A. Weaver}
\affil{Johns Hopkins University Applied Physics Laboratory, Laurel, MD, USA}

\author[0000-0002-5846-716X]{Catherine B. Olkin}
\author[0000-0002-3672-0603]{Joel W. Parker}
\author{S. Alan Stern}
\affil{Southwest Research Institute, 1050 Walnut St. Suite 300, Boulder, CO 80302, USA}

\begin{abstract}
    NASA's New Horizons spacecraft will conduct a close flyby of the cold classical Kuiper Belt Object (KBO) 
    designated (486958) 2014 MU69 on January 1, 2019.
    At a heliocentric distance of 44 AU, "MU69" will be the most distant object ever visited by a spacecraft.
    To enable this flyby, we have developed an extremely high precision orbit fitting and uncertainty 
    processing pipeline, making maximal use of the Hubble Space Telescope's Wide Field Camera 3 (WFC3) 
    and pre-release versions of the ESA Gaia Data Release 2 (DR2) catalog.
    This pipeline also enabled successful predictions of a stellar occultation by MU69 in July 2017.
    We describe how we process the WFC3 images to match the Gaia DR2 catalog, extract positional 
    uncertainties for this extremely faint target (typically 140 photons per WFC3 exposure), 
    and translate those uncertainties into probability distribution functions for MU69 at any given time.
    We also describe how we use these uncertainties to guide New Horizons, plan 
    stellar occultions of MU69, and derive MU69's orbital evolution and long-term stability.
\end{abstract}


\section{Introduction} \label{sec:intro}

The cold classical Kuiper Belt Object (KBO) (486958) 2014 MU$_{69}$ is the primary target for NASA's 
\textit{New Horizons} Kuiper Belt Extended Mission.
The cold classical Kuiper Belt consists of objects on low-eccentricity, 
low-inclination ($<5^\circ$ to the invariant plane) orbits (that is, dynamically ``cold'') 
with heliocentric semimajor axes between about 40 and 50 AU.
The cold classical objects were likely formed in-place and escaped perturbation 
from their initial orbits by giant planet migration \citep[][and references therein]{2011ApJ...738...13B}, 
making them the most distant known remnants of the original protoplanetary disk.

NASA's \textit{New Horizons} spacecraft was launched January 19, 2006, 
received a gravitational assist from Jupiter on February 28, 2007,
and flew through the Pluto-Charon system on July 14, 2015
\citep{2015Sci...350.1815S}.
Since the Pluto encounter, \textit{New Horizons} has observed the 3:2 Neptune resonant (15810) Arawn 
(provisionally designated 1994 JR$_{1}$) in 2016 as close at a distance of 0.7 AU \citep{2016ApJ...828L..15P}.
\textit{New Horizons} will encounter many other KBOs within 1 AU, some as close as 0.1 AU,
and some (such as Quaoar and Haumea) that are much farther away,
but all can be seen by \textit{New  Horizons} at much higher solar phase angles
than is possible from Earth-based telescopes (Porter et al. 2018 in preparation, Verbiscer et al. 2018 in preparation).
However, none of these KBOs will be seen as close as MU$_{69}$, which will be 
within 3500 km of the spacecraft on the nominal trajectory.
\textit{New Horizons} will image the surface of MU$_{69}$ at best resolutions of $\approx$35 meter/pixel,
and spectral maps at $\approx$1 km/pixel.
In order to guide the spacecraft to such a close encounter with a KBO that only has a relatively short orbital arc
required a completely new approach to orbit determination and uncertainty analysis, which we describe in this paper.

2014 MU$_{69}$ was discovered in July 2014 by the \textit{Hubble Space Telescope} (HST) following eight years of 
dedicated searches for a second \textit{New Horizons} encounter object (Buie et al. 2018, in preparation).
After several ground-based searches down to V$\approx$26, 194 HST
orbits were allocated for a deeper, more systematic search for objects accessible 
to \textit{New Horizons} (GO 13633, PI Spencer).
MU$_{69}$ was initially detected in 10 images acquired in two HST orbits,
as were four other KBOs during the \textit{HST} search.
Three objects were potential targets for \textit{New Horizons}: 
2014 MU$_{69}$, 2014 OS$_{393}$, and 2014 PN$_{70}$.
In August 2015, the \textit{New Horizons} team selected 2014 MU$_{69}$ as the potential \textit{New Horizons} extended mission target.
The spacecraft performed a series of four burns in October-November 2015 to target 2014 MU$_{69}$.
The \textit{New Horizons} Kuiper Belt Extended Mission was approved by NASA after Senior Review in July 2016,
and its centerpiece is the flyby of 2014 MU$_{69}$ on January 1, 2019.

In this paper, we will discuss our process of performing absolute astrometry on 2014 MU$_{69}$
tied to a pre-release version of \textit{Gaia} DR2,
propagating that error forward to orbital uncertainty, and then using the orbital uncertainty to
guide both occultations of the KBO and to guide the spacecraft to a close flyby.
These techniques represent the highest-precision heliocentric orbit fitting of a Kuiper Belt object
ever, and can provide a basis for future applications of \textit{Gaia}-driven astrometry to small bodies in the solar system.

\section{Data Sources} \label{sec:data}

2014 MU$_{69}$ was discovered with the \textit{HST} search program (GO 13633) described in (Buie et al. 2018, in preparation).
This program was designed to take five full-frame 370-second Wide Field Camera 3 (WFC3) UVIS images in one orbit with the F350LP 
broadband filter, skip an \textit{HST} orbit, and then repeat the same observation.
The images were tracked on a nominal cold-classical orbit, producing streaked stars,
but non-streaked KBOs.
To perform the search, the images were shift-stacked at 20 different representative cold-classical shift rates.
Objects that appeared in both search orbits for one of the shift rates were identified and targeted for follow up 
\textit{HST} observations.
The first object identified this way was designated 1110113Y (HST orbit IDs 11/12, WFC3 CCD 1, shift rate ID 011, random ID 3Y),
later given the provisional designation 2014 MU$_{69}$.
Four more KBOs were subsequently detected, all of which were brighter, but all of which required more fuel 
for \textit{New Horizons} to reach.

The available dataset for 2014 MU$_{69}$ has both a short temporal arc (July 2014-October 2017),
and an extremely high data quality, making it ideal for the analysis described below.
MU$_{69}$ is a very faint object, with V$\approx$27.5 (Benecchi et al. 2018, in preparation),
and is in an extremely crowded star field (galactic longitudes from -8$^\circ$ to -12$^\circ$).
These constraints have made it effectively impossible to detect with ground-based telescopes,
and all observations of MU$_{69}$ have been with the \textit{Hubble Space Telescope}.
A list of these \textit{HST} observations is in Table \ref{tab:astro}.

Both the initial follow up observations and most observations conducted since then have adopted the search program's basic
format of five 367 to 370-second, F350LP filter, full frame UVIS WFC3 images.
The key exceptions are the color campaign in summer of 2016 (GO 14092, PI Benecchi),
in which four orbits each included two 348-second images using F606W followed by three 373-second images with F814W.
As shown in Table \ref{tab:astro}, 
half of the follow up orbits were roughly evenly spread over August 2014-October 2017,
while the other half were spread over a roughly one week interval in June-July 2017.
The latter was the lightcurve campaign (GO 14627, PI Benecchi), which was critical in successfully predicting the July 2017 occultation
(See Section \ref{sec:occult}).

In addition to \textit{HST} and the July 2017 occultation, the other data source for this analysis was stellar astrometry from the ESA \textit{Gaia} project.
Initially, the process in Section \ref{sec:image} was built using a custom star catalog built from
a deep composite of the MU$_{69}$ field obtained using the Canada-France-Hawaii Telescope \citep{2014JInst...9C4003G}.
When the \textit{Gaia} Data Release 1 (DR1) was made available in September 2016 \citep{2016A&A...595A...2G}, we began using that,
applying a mean proper motion correction to images significantly after the DR1 2015.0 epoch.
The mean proper motion was calculated from the \textit{Gaia} TGAS catalog \citep{2015A&A...574A.115M}.
The MU$_{69}$ observation fields were in areas of very low coverage for TGAS, so the TGAS stars could not be used directly.
With the success of our application of DR1 and after a special support request to the \textit{Gaia} project, 
we were able to obtain a sky patch from a pre-release version of Data Release 2 (DR2) around the path of MU$_{69}$.
The major advances with DR2 are proper motion for all catalog stars, obviating the need for a mean
proper motion correction, as well as a much more homogeneous, bias-free distribution of errors on the sky, 
and much lower uncertainty (one order of magnitude).
This early version of DR2 was used to plan the three 2017 occultations, both for correcting the \textit{HST} absolute
astrometry and for knowledge of the occultation stars themselves.
We obtained a second preview version of DR2 in our field of interest in October 2017, and that version
is used for the astrometry in Table \ref{tab:astro}.

\section{Image Analysis} \label{sec:image}

\begin{figure*}[t]
\plotone{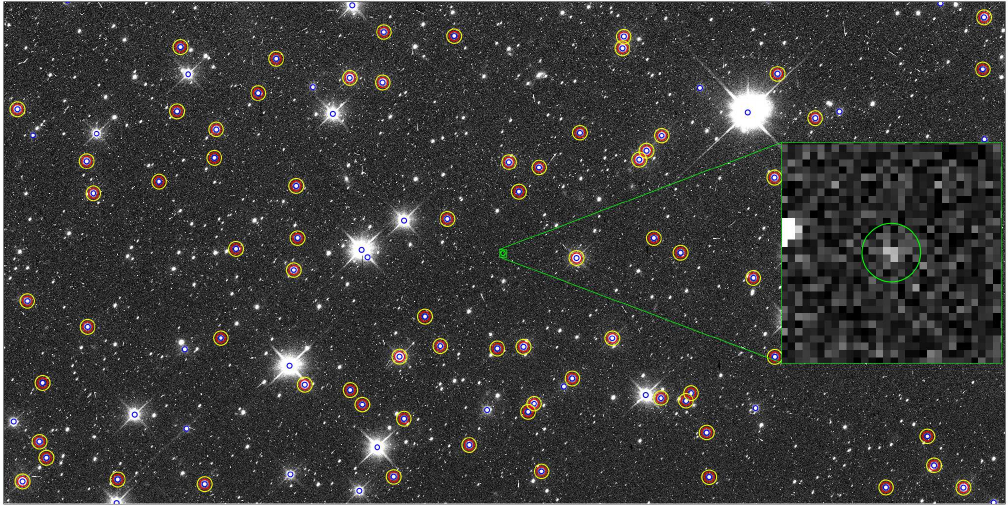}
\caption{An image from the HST lightcurve campaign (id8i16moq).
The blue circles show the \textit{Gaia} DR2 positions of stars,
the red circles show the locations of star PDFs used for the WCS solution,
and the outer yellow circles indicate a match 
(all PDF stars were matched in this case, which is typical).
The inset green box shows the location of 2014 MU$_{69}$;
see Figure \ref{fig:corner}.
\label{fig:hst}}
\end{figure*}

A typical \textit{HST} MU$_{69}$ image is shown in Figure \ref{fig:hst}.
Over the course of 2014-2017, MU$_{69}$ has moved between galactic latitudes of $-8^\circ$ and $-12^\circ$.
Accordingly, the background star density has always been very high in any images of MU$_{69}$ 
(from both Earth and \textit{New Horizons}), 
and this dense background field will persist through the \textit{New Horizons} encounter.
To mitigate this background, we developed a Python program called \textit{warpy.py} to perform simple star subtraction.
Because almost every observation sequence consists of five images of roughly the same field,
\textit{warpy.py} iterates though the images, warps the other four images of the visit to the frame of the fifth,
median combines the four warped images, and subtracts them from the target image.
The images are coregistered by matching sources detected with \textit{Source Extractor} \citep{1996A&AS..117..393B}.
Because the stars in the images are all smeared differently, they do not subtract cleanly.
However, the star-subtracted images are useful for verification that the following steps are fitting 
the KBO and not a background source.

The smeared stars are also a problem for determining the pointing of the images.
Since \textit{HST} is tracking on the motion of the object, a single Tiny Tim \citep{2011SPIE.8127E..0JK}
point spread function (PSF) would not accurately describe the effective PSFs of the stars.
We thus built up "smear kernels" that describe the motion of the stars relative to 
the KBO through the 370 second exposures.
By shifting a Tiny Tim PSF to 400 discrete times during the exposure and averaging them,
we were able to build up an exact model of each star's PSF.
We did this for each star used for the WCS solution (effectively, all the \textit{Gaia} stars in the field),
since the Tiny Tim PSF varies across the WFC3 field.

We next used the stellar PSFs to build up probability distribution functions (PDFs) for the pixel location 
of each star within each image.
We did this task with a Markov-Chain Monte Carlo (MCMC) algorithm implemented in IDL.
This minimal MCMC with a single ``walker'' was iterated for 2000 steps to build a PDF of both pixel position and 
total DN/second flux.
The result is 2000 equal-probability pixel positions for the star, encompassing the true shape of the 
uncertainty distribution.
We found this number of steps to be sufficient, as all the stars used in the astrometric solution had a high 
signal-to-noise ratio and typically there were 70-100 stars used in a given solution.
The flux number was not used directly in the fits, but provided diagnostics if the fits were successful.
A typical star had a 1-$\sigma$ pixel position uncertainty of $<$0.1 pixels, equating to an angular uncertainty of 
$<$4 milliarcseconds.

With these stellar PDFs, we could now build PDFs for the pointing of \textit{HST} in each image.
WFC3 UVIS images use a three-layer World Coordinate System (WCS) to translate pixel to sky coordinates,
as described in \citet{2006A&A...446..747G}.
The first layer is the basic pointing, roll, and trapezoidal warp, the second is a set of SIP polynomials that
describe the low-frequency distortion on the chip, and the third layer is a look-up table that describes
high-frequency pixel distortion (e.g. by irregularities in the lithography of the CCD).
All of the distortion parameters are highly-calibrated, and we only needed to update the pointing 
with deltas to the \textit{CRVAL1} and \textit{CRVAL2} keywords, and the roll by multiplying the CD matrix by a rotation matrix.
Thus for each image we have three parameters for the WCS PDF: delta RA, delta Dec, and delta roll.
We built this WCS PDF by selecting a pixel coordinate for each star from their PDFs, fitting a best-fit
WCS solution, and recording the deltas from the original WCS.
This process was repeated 10,000 times to build a discretely-sampled PDF of the WCS offsets.
The resulting typical 1-$\sigma$ uncertainty in the \textit{CRVAL1} and \textit{CRVAL2} keywords 
(and thus in the pointing of \textit{HST}) was $<$2 milliarcseconds.

The pixel position PDFs for the KBO used the same basic MCMC algorithm as the star pixel PDFs, 
but with the single walker iterated 10000 times to build the PDF.
Since \textit{HST} tracked on the KBO, we did not need a smear kernel to fit the KBO and could use a Tiny Tim PSF directly.
In addition, all but the discovery observations had MU$_{69}$ near the center of WFC3 chip 2 (FITS extension 1),
making for a less distorted PSF than near the chip edges.
Initially, the KBO fit was started with a manual click on the rough position of the KBO in the image.
However, as the orbit improved, this manual position was replaced with a calculated initial position from prior orbit solutions and the WCS.
We also made manual masks of stars and cosmic rays near the KBO that might adversely affect the PDF generation.

\begin{figure}[t]
\plotone{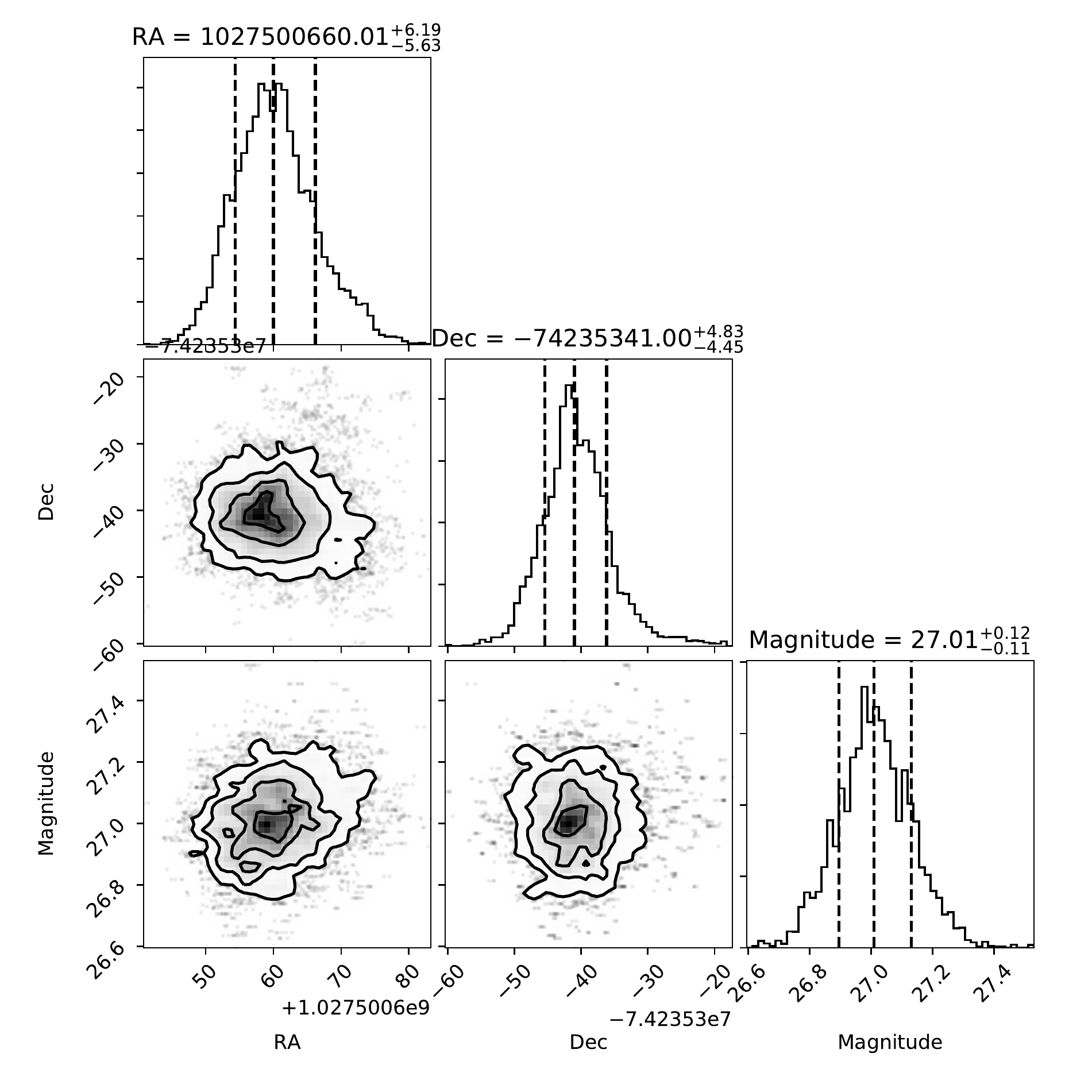}
\caption{The RA/Dec/Magnitude Probability Distribution Function (PDF) for MU$_{69}$
in the image id8i16moq (see Figure \ref{fig:hst}).
RA and Dec are in milliarcseconds, so as to show the uncertainty in appropriate units.
We generated similar PDFs for all the astrometry shown in Table \ref{tab:astro}.
\label{fig:corner}}
\end{figure}

Finally, we needed to combine the KBO pixel PDFs with the WCS PDFs to make KBO sky PDFs. 
Similarly to the WCS PDFs, this was accomplished by selecting a randomly-selected KBO pixel location with a randomly-selected 
WCS PDF, translating to RA and Dec, and then repeating 10,000 times.
An example of one of these PDFs is shown in Figure \ref{fig:corner}.
The instrument magnitude in the pixel PDF was converted to AB apparent magnitude with the PHOTFLAM header keyword.
While the magnitude was not used directly for astrometry, it was an important diagnostic for the quality of the pixel PDF.
For the solutions presented, we filtered out any points with magnitude uncertainties larger than 0.5,
which were generally failures to fit the object,
or any points with uncertainties smaller than 0.1 magnitudes,
typically a cosmic ray close to the KBO causing spuriously high signal-to-noise ratios.
We also only used the F350LP points for the orbit, as the narrower-band points had worse signal-to-noise ratios for both stars and MU$_{69}$.
This left 214 of the 264 images, which we used for an initial fit.
An additional nine points were rejected because they had greater than 30 milliarcsecond residuals relative to the initial fit.
Our final HST astrometry thus used 205 of the 264 images (78\%); these are shown in Table \ref{tab:astro}.

After the method described here was used to successfully predict the occultation of MU$_{69}$ on July 17, 2017
(See Section \ref{sec:occult}, Buie et al. 2018, in preparation),
we were able to use the occultation itself as a high-quality occultation point.
Because five solid-body chords were obtained on July 17, we chose the mid-time of the longest chord
and used it as the nominal center-of-figure.
We could then combine this mid-time, the topocentric location of the portable telescope that obtained the longest chord, 
and the location and uncertainty of the occultation star from \textit{Gaia} DR2 to produce an effective
astrometric PDF.
See Buie et al. (2018, in preparation) for more details about the circumstances and analysis of this occultation.
This occultation PDF could then be combined with the \textit{HST}-derived PDFs in the process described in Section \ref{sec:orbit}.

\section{Orbit Determination} \label{sec:orbit}

Typically, small body orbits in the literature are described in either mean or osculating heliocentric elements,
with error bars representing a normal error distribution.
This is typically sufficient for general dynamical studies and rough targeting from the ground, but not
for spacecraft flybys or occultation planning.
The actual uncertainty of an object's astrometry is rarely perfectly described by a normal distribution,
and neither is that object's location and velocity in space.
We thus sought to develop an orbit-fitting method that would accurately map 
the full astrometric uncertainty distribution into the ephemeris.

To perform these fits, we developed a high-precision few-body orbital integrator.
Since 2014 MU$_{69}$ is a cold-classical KBO, all of the planetary perturbations on it are interior, 
and tend to result in a slow precession.
Non-gravitational factors (i.e. YORP) and general relativity are not a factor for Kuiper Belt objects.
We therefore developed a few-body conservative force integrator, capable of modeling 
the major planets and their perturbing forces on a massless test particle.
This integrator (PyNBody\footnote{\url{https://github.com/ascendingnode/PyNBody}})
is based on the 12/13th order Runge-Kutta-Nystrom intergrator of \citet{Brankin_1989} 
and was previously described in \citet{2016ApJ...828L..15P}.
This integrator is not the fastest, but it is very accurate and can typically conserve 
system energy and momentum to within machine precision over the relevant timescales for orbit fitting.

The KBO's orbit is parameterized as a cartesian state vector relative to the solar system barycenter at a fixed epoch.
The inertial frame for the integrations the International Celestial Reference Frame (ICRF).
\textit{Gaia} DR1 is aligned to ICRF by matching optical detections of quasars with a subset of ICRF2,
while \textit{Gaia} DR2 uses several thousand quasars from ICRF3 and half a million AGNs to perform frame alignment 
\citep{2016A&A...595A...2G}.
The integration epoch is set to 2014-06-01 00:00:00.000 UTC, a few weeks before the first observation 
(originally a safety factor in case of any precoveries in the \textit{HST} search).
To test the solution against the data, we propagate the state vector with the PyNBody integrator to the desired time
and calculate its apparent ICRF RA/Dec from \textit{HST}, with appropriate light-time correction.
We use the JPL NAIF \textit{HST} and \textit{DE430} SPICE kernels to determine the location of \textit{HST} 
relative to the solar system barycenter \citep{2014IPNPR.196C...1F}.

We used the \textit{emcee} Markov-Chain Monte Carlo package \citep{2013PASP..125..306F}
to translate the astrometric uncertainty to orbital uncertainty.
The \textit{emcee} package provides a fast and natively multithreaded way to run MCMC from Python.
As input to \textit{emcee}, the fitting program calculates
the likelihood for any solution by taking the predicted RA/Decs for that solution and comparing them to the RA/Dec PDFs.
Because the PDFs are discretely-sampled, we created a Kernel Density Estimator (KDE) for each observations, using
Silverman's Rule of Thumb \citep{1986desd.book.....S} to choose the bandwidths, since most of the PDFs were roughly Gaussian.
The log likelihoods for all the images could then be summed to provide a total log likelihood to \textit{emcee}.

For any solution, the first step is to make an initial guess (typically an older solution) and minimize its $\chi^2$ with 
a downhill simplex method \citep{doi:10.1093/comjnl/7.4.308}.
This polished solution is then used to create 200 slightly perturbed state vectors as the initial ``walkers'' for
\textit{emcee} to use.
We then run the 200 walkers for 100 iterations to ``burn-in'' and allow them to move away from the artificial initial
distribution.
We then reset \textit{emcee} and run it for 500 iterations to produce the full PDF cloud of 10,000 state vectors at the 
fitting epoch.
These numbers of iterations were arrived at after much testing, and are typically more burn-in than is actually 
necessary, so as to ensure that the solutions are well-distributed.
We save the resulting state vector PDF in a format that can then be propagated to any time of interest.
The state vector and orbit for our ``rd2b'' orbit solution are presented in Table \ref{tab:state}

\begin{deluxetable}{ccccc}[t]
\tablecaption{ The ``rd2b'' orbit solution for 2014 MU$_{69}$.
State vector and orbit are relative to the solar system barycenter
and in the ICRF Ecliptic frame at the epoch 2014-06-01 00:00:00 UTC.
\label{tab:state}}
\tablehead{\colhead{} & \colhead{Value} & \colhead{} & \colhead{1-$\sigma$} }
\startdata
$x$  & $+1.163133074444e+09$ &$\pm$& $2.80233e+02$ & km \\
$y$  & $-6.385039581373e+09$ &$\pm$& $1.52754e+03$ & km \\
$z$  & $+2.373261916929e+08$ &$\pm$& $5.87015e+01$ & km \\
$v_x$ & $+4.461378977476e+00$ &$\pm$& $5.92714e-06$ & km/s \\
$v_y$ & $+9.619622770583e-01$ &$\pm$& $2.45488e-05$ & km/s \\
$v_z$ & $-1.066958207821e-01$ &$\pm$& $9.45150e-07$ & km/s \\
\hline
$a$ &  44.23555350  &$\pm$&  0.00003999 & AU \\
$e$ &   0.03787388 &$\pm$&   0.00000476 & \\
$I$ &   2.44993086 &$\pm$&   0.00000203 & deg \\
$\Omega$ & 159.04712465 &$\pm$&   0.00006746 & deg \\
$\omega$ & 183.74800591 &$\pm$&   0.00469779 & deg \\
$M$ & 301.30454775 &$\pm$&   0.00406885 & deg \\
\enddata
\end{deluxetable}

While the full PDF cloud of 10,000 states encapsulates the uncertainty in the location of MU$_{69}$ any any time, 
it is rather unwieldy to use in most circumstances.
We therefore calculated the state clouds of 2014 MU$_{69}$ at 2000 discrete times between January 1, 2004 and
January 1, 2024, and averaged the states at each time.
We then generated order-27 Chebyshev polynominals \citep{tchebychev1853theorie} for the positions of the KBO, and saved them in a 
JPL SPICE Type 02 SPK kernel.
JPL SPICE could then be used to rapidly interpolate the location of MU$_{69}$ at any time along the interval.
Testing the kernel at 769 random points over the interval returned a root-mean-squared residual of 20 meters,
well below the uncertainty in the orbit.

\section{New Horizons Trajectory Planning} \label{sec:traj}

The primary reason to determine the orbit of 2014 MU$_{69}$ to very high precision is to ensure the success of 
the \textit{New Horizons} flyby.
\textit{New Horizons} performed the major Trajectory Correction Maneuver (TCM) to guide it to MU$_{69}$ over a series 
of four burn segments in October and November 2015, after all Pluto observations had finished.
The initial orbit used to target the spacecraft was based on the first year of data, from June 2014 to July 2015
(GO 13633 and GO/DD 14053, PI Spencer).
After that burn, early versions of the analysis described here showed that significantly more \textit{HST} observations 
would be required to enable a close flyby of 2014 MU$_{69}$.
We thus proposed and were awarded six \textit{HST} orbits in 2016, and five in 2017 (GO/DD 14485, GO 14629, and GO 15158, PI Buie).
In addition, 24 \textit{HST} orbits were used in June/July 2017 to measure the lightcurve of MU$_{69}$ (GO 14627, PI Benecchi).
The orbit presented here uses data from all of these \textit{HST} programs, in addition to the July 17 occultation.

The \textit{New Horizons} spacecraft will nominally fly closest to 2014 MU$_{69}$ at 05:33 January 1, 2019 UTC.
This time was chosen to enable both the Goldstone and Canberra Deep Space Network (DSN) 70-meter dishes to uplink to the
spacecraft simultaneously for an attempted bistatic radar experiment 
\citep[as was performed at Pluto;][]{2016DPS....4821304L}.
\textit{New Horizons} will not be able to acquire MU$_{69}$ any earlier than August 2018.
Because of \textit{New Horizons}'s almost radial trajectory out of the solar system, the KBO will move very slowly 
against the background stars until a just few weeks before encounter.
While the spacecraft can use the LOng Range Reconnaissance Imager \citep[LORRI,][]{2005SPIE.5906..407C} 
to well constrain the location of 2014 MU$_{69}$ in the ``B-Plane''
(the plane perpendicular to the spacecraft's motion and containing the flyby target), 
the time-of-flight (ToF) uncertainty along the direction of the 
spacecraft's motion is constrained only by the Earth-based orbital solution.
The distance to the spacecraft from Earth will be well-constrained by Doppler radio measurements on approach to MU$_{69}$,
and so the uncertainty in the absolute location of MU$_{69}$ relative to the solar system barycenter will 
determine the flyby ToF uncertainty.

\section{Application to Occultation Planning} \label{sec:occult}

In addition to guiding \textit{New Horizons}, we also used our orbit solution to predict three stellar occultations
by 2014 MU$_{69}$ in 2017 and one in 2018.
The 2017 occultation campaign is comprehensively described in Buie et al. (2018, in preparation) and here we detail only the procedures 
used to predict the occultations.
MU$_{69}$ is a small object, with an absolute magnitude of $H_V\approx11$ Benecchi et al. (2018, in preparation), 
corresponding to a size likely smaller than 50 km diameter.
We therefore knew that the occultations would only be successful if we had very high-quality orbital estimates 
and uncertainty models.
Thankfully, that is exactly what we had developed for guiding \textit{New Horizons}.

Stellar occultations occur when a solar system object passes in front of a star from the perspective of an observer.
They have been used to discover the atmosphere of Pluto \citep{1989Icar...77..148E}
and the rings around Uranus, Chariklo, and Haumea 
\citep[respectively]{1977Natur.267..328E,2014Natur.508...72B,2017Natur.550..219O}.
The latter is most important for planning the \textit{New Horizons} flyby of MU$_{69}$, as occultations
provide the only way of detecting rings or other opacity structures around the KBO before the spacecraft 
is close enough to see them directly.
In addition, occultations can (and in this case did) provide estimates of the size and shape of a body.
Knowledge of the approximate size of MU$_{69}$ enabled estimates of its bulk albedo, 
and therefore allowed mission planners to better estimate the correct exposure times for the flyby images.

Because of the motion and rotation of the Earth, stellar occultations sweep across Earth from west to east.
The typical approach to observe an occultation of an object with an uncertain orbit is therefore 
to set up a north-south ``picket fence'' of portable telescope teams perpendicular or ``crosstrack'' to the occultation path.
It is therefore important to know both the crosstrack uncertainty of the prediction, and what the crosstrack 
distance of any station is.
The latter often requires some iteration, as finding a logistically viable site with the proper crosstrack 
can be challenging, especially in an unfamiliar country.
We thus developed tools to export KML files to Google Maps with lines showing the target crosstracks for each observing team.
These could be used for planning site reconnaissance, and with GPS-enabled smartphones, used to see in real time where
a potential site was located compared to the desired crosstrack line.
To estimate the crosstrack uncertainty, we propagated the full 10,000 states to the occultation time and 
calculated the geometry for all of those states.
This produced a full PDF of the occultation uncertainty, which we could use to plan the crosstracks to maximize the
chance of success.

The first MU$_{69}$ occultation of 2017 was on June 3.
The ground track for this event passed over both South America and South Africa,
and 25 portable telescope teams were deployed to both Mendoza, Argentina and 
the Northern Cape and Western Cape Provinces of South Africa.
Two \textit{HST} astrometric observations of MU$_{69}$ had been planned for the spring of 2017,
on March 16 and in mid May.
However, the March observation failed due to a safing event on \textit{HST}, 
and the observation could not be rescheduled in April,
because MU$_{69}$ was passing through quadrature and did not have enough sky motion 
from \textit{HST} to be detected.
Thus, the first MU$_{69}$ observations of 2017 were acquired by \textit{HST} on May 1 and 25.
An initial solution though May 1, ``may1c'', was used to plan the deployments to Argentina and South Africa.
This was superseded by the ``may25a'' solution with data through May 25,
which was produced after the orbit fitters (Porter and Buie) had deployed, 
and was used to plan the actual ground tracks.
The ``may25a'' solution purported to have a crosstrack uncertainty of 44 km,
though the subsequent June-July 2017 observations showed that the ``may25a'' ground track
had been too far north by almost 2-sigma.
This offset precluded a solid-body occultation on June 3, though the high signal-to-noise ratio observations
of the event at the South African Astronomical Observatory 74-inch telescope and 
at Gemini South did exclude optically-thick rings around MU$_{69}$.
See Buie et al. (2018, in preparation) for more details.

\begin{deluxetable}{ccc}[t]
\tablecaption{ Mean, Free and Forced Elements for Best-Fit MU$_{69}$.
\label{tab:mean}}
\tablehead{\colhead{Best Fit,} & \colhead{10$^8$ years} }
\startdata
Mean $a$ & 44.23 AU \\
Forced $e$ & 6 x 10$^{-5}$ \\
Free $e$ & 0.037 \\
Forced $i_{mss}$ & 0.26$^\circ$ \\
Forced $i_{mkb}$ & 0.0012$^\circ$ \\
Free $i$ & 2.54$^\circ$ \\
\enddata
\end{deluxetable}

The second MU$_{69}$ occultation of 2017 was on July 10.
This event occurred mainly over the Pacific Ocean with a much dimmer star (V$\approx$15.5)
and nearly-full Moon, thus preventing a ground-based observation campaign.
However, the NASA-DLR Stratospheric Observatory for Infrared Astronomy (SOFIA) 
airborne observatory was able to reach the occultation track from its
southern deployment base in Christchurch, New Zealand, 
and NASA awarded a flight to observe the July 10 occultation (PI E. Young).
Between the June 3 and July 10 events, 
\textit{HST} observed 2014 MU$_{69}$ over 24 orbits between June 25 and July 4 (GO 14627, PI Benecchi).
This program provided a wealth of new images to integrate into the MU$_{69}$ orbit solution
in a very short amount of time,
and time was especially critical, as the last six orbits worth of data was downlinked from \textit{HST}
after the orbit fitting team (Porter and Buie) had arrived in Christchurch.
Thus, the orbit solution used to guide the SOFIA flight was necessarily determined at the 
United States Antarctic Program Christchurch facility, and delivered to the SOFIA mission
planners 36 hours before the flight.
This orbit solution, ``lc1'', used all of the lightcurve campaign points, 
plus the highest-quality preceding \textit{HST} MU$_{69}$ observations.

The final MU$_{69}$ occultation of 2017 on July 17 was observed 
with portable ground stations in the Chubut and Santa Cruz provinces in Argentina.
No additional observations of \textit{HST} MU$_{69}$ were made between the July 10 and July 17 occultations,
but we did perform a more thorough filtering of low-quality points.
This resulted in the ``lc1gr'' solution that was used to guide the placement of stations for 
the July 17 occultation.
The ``lc1gr'' solution had a 1-$\sigma$ uncertainty in crosstrack of 13 km, much tighter than for June 3.
This solution allowed a tighter picket fence of stations up and down the Patagonian coast, centered
a few kilometers north of the city of Comodoro Rivadavia.
Despite high winds on the occultation night, 22 of the 24 deployed stations successfully observed the occultation.
Five of those stations observed the solid-body occultation, with the southernmost being the
predicted centerline from the ``lc1gr'' solution.

This work predicts
an additional stellar occultation opportunity on August 4, 2018.
The ground track for this event passes over 
western Africa (Mail, Mauritania, and Senegal)
and northern South America (Guyana, Venezuela, and Colombia).
With the ``rd2b'' solution presented in Table \ref{tab:state}, 
the 1-$\sigma$ crosstrack uncertainty is 12 km.
This uncertainty will decrease somewhat with additional \textit{HST} observations in 2018.

\vspace{12pt}
\section{Long-Term Orbital Evolution} \label{sec:longterm}

\begin{figure}[t]
\plotone{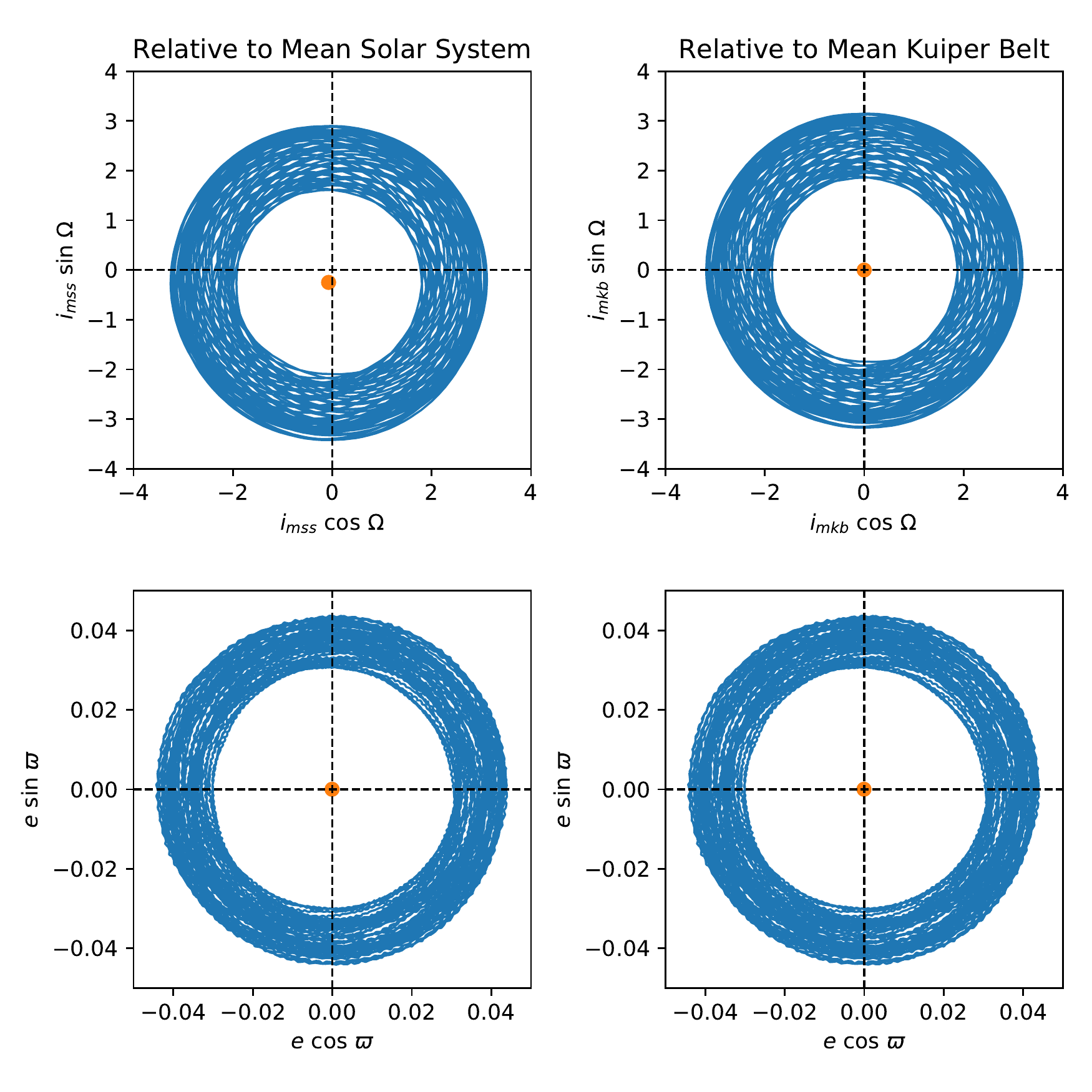}
\caption{Free and forced elements for best-fit 2014 MU$_{69}$;
the centers of the circles show forced inclination/eccentricity, 
while the radii show free inclination/eccentricity.
The forced inclination is 0.26$^\circ$ from the mean solar system 
angular momentum vector (left),
but almost perfectly fits the mean Kuiper Belt at 44 AU pole from
\citet[right]{2017AJ....154...62V}.
The lack of forced eccentricity or inclination implies that MU$_{69}$
has not experienced any significant orbital evolution since formation.
\label{fig:ei}}
\end{figure}

2014 MU$_{69}$ is a cold classical Kuiper Belt object.
\citet{2005AJ....129.1117E} identified the ``Classical'' KBOs as non-resonant objects 
with eccentricities smaller than 0.2.
This classification was further refined as a ``Cold Classical'' or ``Kernel'' 
population by \citet{2011AJ....142..131P}
with $a\approx44$ AU, $e\approx0.05$, $i<5^\circ$ to the invariant plane.
MU$_{69}$ has $a$=44.2 AU, $e$=0.03, $i$=2.4$^\circ$,
making it an archetype of the cold classical population.
\citet{2011ApJ...738...13B} showed the orbits of cold classical objects were likely formed 
in-place and survived being disturbed from their initial orbits by giant planet migration.
The unusually high binary fraction of cold classical KBOs \citep{2008ssbn.book..345N}
is an additional line of evidence that they are mostly undisturbed from their original orbits.
Indeed, the observed cold classical KBO binary fraction is high enough that nearly all must have 
originally formed as binaries or higher-order multiple systems \citep{2017NatAs...1E..88F}.

With a few small modifications to the PyNBody code, we were able to integrate the orbit of MU$_{69}$ over 
sufficiently long timescales to test this stability and determine mean orbital elements.
Specifically, we changed the time unit in the integration from seconds to years to allow for longer 
integrations without worry of overflows and removed the terrestrial planets as perturbers 
(instead dropping their masses into the Sun).
The results of integrations forward and back $10^8$ years can be seen in Figure \ref{fig:ei} and Table \ref{tab:mean},
projected in both the mean solar system plane defined by the \textit{de430.bsp} planets in the ICRF J2000 Ecliptic frame, 
$i_m=1.6^\circ$ and $\Omega_m=72.4^\circ$,
and in the mean Kuiper Belt plane at 44 AU as determined from known KBOs
\citet{2017AJ....154...62V}, $i_m=1.8^\circ$ and $\Omega_m=77.0^\circ$.
The mean, free, and forced elements of MU$_{69}$'s orbit are shown in Table \ref{tab:mean}.
The forced inclination of MU$_{69}$ to the mean solar system plane is 0.26$^\circ$, 
but only 0.0012$^\circ$ to the mean Kuiper Belt at 44 AU.
Likewise, the forced eccentricity of MU$_{69}$ is less than 0.0001.
The apparent lack of any forced inclination or eccentricity to the mean Kuiper Belt is strong evidence
that MU$_{69}$ has not suffered any significant orbital evolution beyond secular perturbations.
MU$_{69}$ should therefore represent a truly pristine fossil of the Sun's protoplanetary disk,
an object unlike any other previously  visited by a spacecraft.

\section{Summary} \label{sec:summary}

We have described the process we have used to fit the orbit of
2014 MU$_{69}$, as of the start of 2018.
This process combines \textit{Gaia} DR2 and \textit{HST}/WFC3 to produce extremely high
precision absolute astrometry of MU$_{69}$,
and translates that uncertainty into a cartesian state vector 
probability distribution function that can be evolved
to any time of interest.
The results of this analysis were used to successfully predict and observe a solid-body stellar 
occultation of MU$_{69}$ on July 17, 2017, predict a stellar occultation on August 4, 2018,
and to guide the \textit{New Horizons} spacecraft
to a close (3500 km) flyby of MU$_{69}$ on January 1, 2019.

The process described here should enable high-precision orbit determination
for future occultations and spacecraft missions.
2014 MU$_{69}$ presents the extreme case of a very interesting object
that is both faint and in a very crowded star field.
Now that the \textit{Gaia} DR2 catalog has been released,
solar system objects with higher signal-to-noise ratios should benefit 
even more from this technique, 
enabling a substantial improvement in orbital uncertainty and 
increasing the number of objects that might be observed with stellar 
occultations.

\acknowledgments
This work was supported by NASA's \textit{New Horizons} mission 
and \textit{HST} programs GO 13633, GO/DD 14053, GO 14092, GO/DD 14485, 
GO 14627, GO 14629, and GO 15158.
Support for this program was provided by NASA through a grant from the Space Telescope Science Institute, 
which is operated by the Association of Universities for Research in Astronomy, Inc., under NASA contract NAS 5-26555.
Special thanks to ESA for providing pre-release versions of \textit{Gaia} DR2
over the relevant regions.
Special thanks to Bill Folkner for providing his expertise and verification of the orbit fitting results.

\facility{HST(WFC3), Gaia, SOFIA}

\software{
	astropy \citep{2018arXiv180102634T},
    scipy \citep{Scipy},
    emcee \citep{2013PASP..125..306F},
    Matplotlib \citep{Hunter:2007},
    photutils,
    spiceypy
}

\bibliography{refs1}

\appendix

\startlongtable
\begin{deluxetable*}{ccccrr}
\tablecaption{ HST/WFC3 astrometry for MU$_{69}$.
\label{tab:astro}}
\tablehead{\colhead{WFC3} & \colhead{} & \colhead{} & 
\colhead{} & \colhead{$\sigma$ RA} & \colhead{$\sigma$ Dec} \\
\colhead{Dataset} & \colhead{UTC Time} & \colhead{RA} & 
\colhead{Dec} & \colhead{(mas)} & \colhead{(mas)}}
\startdata
icii11r7q & 2014-06-26T08:51:42.4042 & $18^\textrm{h}45^\textrm{m}10\overset{\textrm{s}}{.}67179$ & $-20^\circ53^\prime03\overset{\prime\prime}{.}0565$ &  6.496 &  5.874 \\
icii11r8q & 2014-06-26T09:00:33.4117 & $18^\textrm{h}45^\textrm{m}10\overset{\textrm{s}}{.}63684$ & $-20^\circ53^\prime03\overset{\prime\prime}{.}1070$ &  8.280 &  8.519 \\
icii11raq & 2014-06-26T09:09:24.4036 & $18^\textrm{h}45^\textrm{m}10\overset{\textrm{s}}{.}59521$ & $-20^\circ53^\prime03\overset{\prime\prime}{.}1753$ &  9.150 &  9.504 \\
icii11rcq & 2014-06-26T09:18:15.3964 & $18^\textrm{h}45^\textrm{m}10\overset{\textrm{s}}{.}55800$ & $-20^\circ53^\prime03\overset{\prime\prime}{.}2431$ &  7.130 &  6.348 \\
icii11req & 2014-06-26T09:27:06.4048 & $18^\textrm{h}45^\textrm{m}10\overset{\textrm{s}}{.}52223$ & $-20^\circ53^\prime03\overset{\prime\prime}{.}3121$ &  6.935 &  5.378 \\
icii12rpq & 2014-06-26T12:02:50.4041 & $18^\textrm{h}45^\textrm{m}10\overset{\textrm{s}}{.}01995$ & $-20^\circ53^\prime03\overset{\prime\prime}{.}7671$ &  5.202 &  5.230 \\
icii12rqq & 2014-06-26T12:11:41.4124 & $18^\textrm{h}45^\textrm{m}09\overset{\textrm{s}}{.}98343$ & $-20^\circ53^\prime03\overset{\prime\prime}{.}8268$ &  5.274 &  6.928 \\
icii12rsq & 2014-06-26T12:20:32.4044 & $18^\textrm{h}45^\textrm{m}09\overset{\textrm{s}}{.}94411$ & $-20^\circ53^\prime03\overset{\prime\prime}{.}8733$ &  8.544 &  8.013 \\
icii12ruq & 2014-06-26T12:29:23.3963 & $18^\textrm{h}45^\textrm{m}09\overset{\textrm{s}}{.}90568$ & $-20^\circ53^\prime03\overset{\prime\prime}{.}9530$ &  7.870 &  7.667 \\
icii12rwq & 2014-06-26T12:38:14.4046 & $18^\textrm{h}45^\textrm{m}09\overset{\textrm{s}}{.}87100$ & $-20^\circ53^\prime04\overset{\prime\prime}{.}0172$ & 10.727 & 11.309 \\
iciig7cwq & 2014-08-02T13:05:20.4622 & $18^\textrm{h}42^\textrm{m}15\overset{\textrm{s}}{.}46771$ & $-20^\circ56^\prime43\overset{\prime\prime}{.}2187$ &  8.676 &  9.408 \\
iciig7cyq & 2014-08-02T13:14:11.4541 & $18^\textrm{h}42^\textrm{m}15\overset{\textrm{s}}{.}43399$ & $-20^\circ56^\prime43\overset{\prime\prime}{.}2084$ &  8.216 &  8.844 \\
iciig7d0q & 2014-08-02T13:23:02.4624 & $18^\textrm{h}42^\textrm{m}15\overset{\textrm{s}}{.}40181$ & $-20^\circ56^\prime43\overset{\prime\prime}{.}1822$ &  6.487 &  6.519 \\
iciig7d2q & 2014-08-02T13:31:53.4699 & $18^\textrm{h}42^\textrm{m}15\overset{\textrm{s}}{.}37313$ & $-20^\circ56^\prime43\overset{\prime\prime}{.}1570$ &  9.745 &  7.306 \\
iciig8kaq & 2014-08-03T17:45:21.4709 & $18^\textrm{h}42^\textrm{m}10\overset{\textrm{s}}{.}58094$ & $-20^\circ56^\prime50\overset{\prime\prime}{.}4392$ &  6.943 &  5.157 \\
iciig8kcq & 2014-08-03T17:54:12.4620 & $18^\textrm{h}42^\textrm{m}10\overset{\textrm{s}}{.}54825$ & $-20^\circ56^\prime50\overset{\prime\prime}{.}4185$ &  7.896 &  8.639 \\
iciig8keq & 2014-08-03T18:03:03.4540 & $18^\textrm{h}42^\textrm{m}10\overset{\textrm{s}}{.}51593$ & $-20^\circ56^\prime50\overset{\prime\prime}{.}3898$ &  8.294 & 12.064 \\
iciig9rvq & 2014-08-21T07:39:21.5326 & $18^\textrm{h}41^\textrm{m}08\overset{\textrm{s}}{.}70535$ & $-20^\circ58^\prime32\overset{\prime\prime}{.}3648$ & 10.552 & 10.039 \\
iciig9rwq & 2014-08-21T07:48:12.5236 & $18^\textrm{h}41^\textrm{m}08\overset{\textrm{s}}{.}68016$ & $-20^\circ58^\prime32\overset{\prime\prime}{.}4070$ &  8.175 &  5.865 \\
iciig9ryq & 2014-08-21T07:57:03.5164 & $18^\textrm{h}41^\textrm{m}08\overset{\textrm{s}}{.}65466$ & $-20^\circ58^\prime32\overset{\prime\prime}{.}5227$ &  9.206 &  9.342 \\
iciig9s0q & 2014-08-21T08:05:54.5239 & $18^\textrm{h}41^\textrm{m}08\overset{\textrm{s}}{.}62880$ & $-20^\circ58^\prime32\overset{\prime\prime}{.}5833$ &  8.610 &  8.553 \\
iciig9s2q & 2014-08-21T08:14:45.5323 & $18^\textrm{h}41^\textrm{m}08\overset{\textrm{s}}{.}60632$ & $-20^\circ58^\prime32\overset{\prime\prime}{.}6306$ &  7.169 &  8.753 \\
iciih0s4q & 2014-08-21T09:14:54.5242 & $18^\textrm{h}41^\textrm{m}08\overset{\textrm{s}}{.}51133$ & $-20^\circ58^\prime32\overset{\prime\prime}{.}7412$ &  9.255 &  8.626 \\
iciih0s5q & 2014-08-21T09:23:45.5317 & $18^\textrm{h}41^\textrm{m}08\overset{\textrm{s}}{.}48692$ & $-20^\circ58^\prime32\overset{\prime\prime}{.}8053$ & 10.174 &  7.858 \\
iciih0s7q & 2014-08-21T09:32:36.5245 & $18^\textrm{h}41^\textrm{m}08\overset{\textrm{s}}{.}46256$ & $-20^\circ58^\prime32\overset{\prime\prime}{.}8851$ & 11.097 &  6.604 \\
iciih0s9q & 2014-08-21T09:41:27.5164 & $18^\textrm{h}41^\textrm{m}08\overset{\textrm{s}}{.}43633$ & $-20^\circ58^\prime32\overset{\prime\prime}{.}9342$ &  7.150 &  7.405 \\
iciih0sbq & 2014-08-21T09:50:18.5248 & $18^\textrm{h}41^\textrm{m}08\overset{\textrm{s}}{.}41206$ & $-20^\circ58^\prime33\overset{\prime\prime}{.}0122$ &  6.504 &  7.669 \\
iciih3byq & 2014-08-23T04:14:48.5240 & $18^\textrm{h}41^\textrm{m}03\overset{\textrm{s}}{.}42093$ & $-20^\circ58^\prime42\overset{\prime\prime}{.}4838$ &  8.822 &  8.230 \\
iciih3bzq & 2014-08-23T04:23:39.5160 & $18^\textrm{h}41^\textrm{m}03\overset{\textrm{s}}{.}39764$ & $-20^\circ58^\prime42\overset{\prime\prime}{.}5345$ &  8.260 &  7.247 \\
iciih3c1q & 2014-08-23T04:32:30.5243 & $18^\textrm{h}41^\textrm{m}03\overset{\textrm{s}}{.}37316$ & $-20^\circ58^\prime42\overset{\prime\prime}{.}6359$ &  8.590 &  8.633 \\
iciih3c3q & 2014-08-23T04:41:21.5318 & $18^\textrm{h}41^\textrm{m}03\overset{\textrm{s}}{.}34782$ & $-20^\circ58^\prime42\overset{\prime\prime}{.}7163$ & 12.208 &  8.579 \\
iciih3c5q & 2014-08-23T04:50:12.5401 & $18^\textrm{h}41^\textrm{m}03\overset{\textrm{s}}{.}32417$ & $-20^\circ58^\prime42\overset{\prime\prime}{.}7810$ &  8.243 &  8.709 \\
iciih4c7q & 2014-08-23T05:50:22.5239 & $18^\textrm{h}41^\textrm{m}03\overset{\textrm{s}}{.}23676$ & $-20^\circ58^\prime42\overset{\prime\prime}{.}8412$ &  7.167 &  7.342 \\
iciih4c8q & 2014-08-23T05:59:13.5323 & $18^\textrm{h}41^\textrm{m}03\overset{\textrm{s}}{.}21355$ & $-20^\circ58^\prime42\overset{\prime\prime}{.}8952$ &  7.158 &  5.986 \\
iciih4caq & 2014-08-23T06:08:04.5242 & $18^\textrm{h}41^\textrm{m}03\overset{\textrm{s}}{.}18844$ & $-20^\circ58^\prime42\overset{\prime\prime}{.}9833$ &  8.322 &  7.320 \\
iciih4ccq & 2014-08-23T06:16:55.5162 & $18^\textrm{h}41^\textrm{m}03\overset{\textrm{s}}{.}16358$ & $-20^\circ58^\prime43\overset{\prime\prime}{.}0496$ &  8.754 & 10.610 \\
iciih4ceq & 2014-08-23T06:25:46.5236 & $18^\textrm{h}41^\textrm{m}03\overset{\textrm{s}}{.}14306$ & $-20^\circ58^\prime43\overset{\prime\prime}{.}1152$ &  8.151 &  7.563 \\
iciij5ydq & 2014-10-15T01:36:27.6413 & $18^\textrm{h}40^\textrm{m}43\overset{\textrm{s}}{.}94965$ & $-21^\circ01^\prime45\overset{\prime\prime}{.}5109$ &  9.931 & 10.104 \\
iciij5yfq & 2014-10-15T01:45:18.6488 & $18^\textrm{h}40^\textrm{m}43\overset{\textrm{s}}{.}95600$ & $-21^\circ01^\prime45\overset{\prime\prime}{.}5515$ & 11.179 & 10.119 \\
iciij5yhq & 2014-10-15T01:54:09.6572 & $18^\textrm{h}40^\textrm{m}43\overset{\textrm{s}}{.}96306$ & $-21^\circ01^\prime45\overset{\prime\prime}{.}5758$ &  9.929 &  8.851 \\
iciij5yjq & 2014-10-15T02:03:00.6655 & $18^\textrm{h}40^\textrm{m}43\overset{\textrm{s}}{.}96693$ & $-21^\circ01^\prime45\overset{\prime\prime}{.}6258$ &  8.356 &  8.839 \\
iciij5yoq & 2014-10-15T02:11:51.6410 & $18^\textrm{h}40^\textrm{m}43\overset{\textrm{s}}{.}97731$ & $-21^\circ01^\prime45\overset{\prime\prime}{.}6487$ &  7.518 &  7.322 \\
iciij6yyq & 2014-10-15T03:29:42.6497 & $18^\textrm{h}40^\textrm{m}44\overset{\textrm{s}}{.}10758$ & $-21^\circ01^\prime45\overset{\prime\prime}{.}6562$ &  9.126 & 10.111 \\
iciij6z2q & 2014-10-15T03:47:24.6647 & $18^\textrm{h}40^\textrm{m}44\overset{\textrm{s}}{.}12125$ & $-21^\circ01^\prime45\overset{\prime\prime}{.}7481$ &  6.790 &  6.469 \\
iciij7etq & 2014-10-16T04:40:32.6328 & $18^\textrm{h}40^\textrm{m}46\overset{\textrm{s}}{.}48368$ & $-21^\circ01^\prime46\overset{\prime\prime}{.}5691$ & 11.437 &  8.316 \\
iciij7eyq & 2014-10-16T05:07:05.6570 & $18^\textrm{h}40^\textrm{m}46\overset{\textrm{s}}{.}50539$ & $-21^\circ01^\prime46\overset{\prime\prime}{.}7101$ &  8.268 &  8.288 \\
iciij9c6q & 2014-10-22T08:44:42.6722 & $18^\textrm{h}41^\textrm{m}02\overset{\textrm{s}}{.}50094$ & $-21^\circ01^\prime49\overset{\prime\prime}{.}7408$ &  9.897 & 11.963 \\
iciij9c7q & 2014-10-22T08:53:33.6805 & $18^\textrm{h}41^\textrm{m}02\overset{\textrm{s}}{.}51372$ & $-21^\circ01^\prime49\overset{\prime\prime}{.}8238$ & 11.269 &  9.985 \\
iciij9c9q & 2014-10-22T09:02:24.6733 & $18^\textrm{h}41^\textrm{m}02\overset{\textrm{s}}{.}52511$ & $-21^\circ01^\prime49\overset{\prime\prime}{.}8756$ &  7.835 &  6.873 \\
iciij9cdq & 2014-10-22T09:20:06.6727 & $18^\textrm{h}41^\textrm{m}02\overset{\textrm{s}}{.}54616$ & $-21^\circ01^\prime49\overset{\prime\prime}{.}9420$ & 11.968 &  9.530 \\
ict101egq & 2015-05-04T16:36:10.0607 & $18^\textrm{h}54^\textrm{m}04\overset{\textrm{s}}{.}85674$ & $-20^\circ45^\prime17\overset{\prime\prime}{.}1327$ &  8.464 &  8.629 \\
ict101eiq & 2015-05-04T16:44:59.0697 & $18^\textrm{h}54^\textrm{m}04\overset{\textrm{s}}{.}83515$ & $-20^\circ45^\prime17\overset{\prime\prime}{.}0929$ &  8.467 &  7.449 \\
ict101ekq & 2015-05-04T16:53:48.0771 & $18^\textrm{h}54^\textrm{m}04\overset{\textrm{s}}{.}81533$ & $-20^\circ45^\prime17\overset{\prime\prime}{.}0358$ &  8.416 &  8.155 \\
ict101emq & 2015-05-04T17:02:37.0688 & $18^\textrm{h}54^\textrm{m}04\overset{\textrm{s}}{.}79577$ & $-20^\circ45^\prime16\overset{\prime\prime}{.}9791$ &  8.591 &  7.354 \\
ict103vvq & 2015-07-04T14:30:23.1969 & $18^\textrm{h}50^\textrm{m}04\overset{\textrm{s}}{.}47370$ & $-20^\circ49^\prime11\overset{\prime\prime}{.}5823$ &  8.552 & 11.691 \\
ict103vwq & 2015-07-04T14:39:13.1961 & $18^\textrm{h}50^\textrm{m}04\overset{\textrm{s}}{.}43802$ & $-20^\circ49^\prime11\overset{\prime\prime}{.}6027$ &  7.657 &  7.162 \\
ict103vyq & 2015-07-04T14:48:03.1962 & $18^\textrm{h}50^\textrm{m}04\overset{\textrm{s}}{.}39866$ & $-20^\circ49^\prime11\overset{\prime\prime}{.}6196$ &  8.622 &  8.340 \\
ict103w0q & 2015-07-04T14:56:53.1963 & $18^\textrm{h}50^\textrm{m}04\overset{\textrm{s}}{.}36010$ & $-20^\circ49^\prime11\overset{\prime\prime}{.}6368$ &  5.515 &  6.545 \\
ict103w2q & 2015-07-04T15:05:43.1963 & $18^\textrm{h}50^\textrm{m}04\overset{\textrm{s}}{.}32408$ & $-20^\circ49^\prime11\overset{\prime\prime}{.}6547$ &  5.728 &  6.567 \\
id3m01hhq & 2016-03-15T00:11:56.2344 & $18^\textrm{h}59^\textrm{m}12\overset{\textrm{s}}{.}34397$ & $-20^\circ42^\prime13\overset{\prime\prime}{.}6731$ & 10.700 &  8.556 \\
id3m01hiq & 2016-03-15T00:20:44.2342 & $18^\textrm{h}59^\textrm{m}12\overset{\textrm{s}}{.}35540$ & $-20^\circ42^\prime13\overset{\prime\prime}{.}6779$ & 11.293 & 10.247 \\
id3m01hlq & 2016-03-15T00:29:32.2350 & $18^\textrm{h}59^\textrm{m}12\overset{\textrm{s}}{.}36343$ & $-20^\circ42^\prime13\overset{\prime\prime}{.}6355$ &  9.891 & 10.908 \\
id3m01hqq & 2016-03-15T00:38:20.2349 & $18^\textrm{h}59^\textrm{m}12\overset{\textrm{s}}{.}37280$ & $-20^\circ42^\prime13\overset{\prime\prime}{.}5403$ &  8.836 &  7.947 \\
id3m01hsq & 2016-03-15T00:47:08.2340 & $18^\textrm{h}59^\textrm{m}12\overset{\textrm{s}}{.}38636$ & $-20^\circ42^\prime13\overset{\prime\prime}{.}4204$ & 10.946 & 10.261 \\
id3m02soq & 2016-05-15T02:19:56.3589 & $18^\textrm{h}59^\textrm{m}07\overset{\textrm{s}}{.}54093$ & $-20^\circ39^\prime50\overset{\prime\prime}{.}5456$ &  6.365 &  6.087 \\
id5901eqq & 2016-07-25T16:49:31.5321 & $18^\textrm{h}53^\textrm{m}53\overset{\textrm{s}}{.}03748$ & $-20^\circ46^\prime32\overset{\prime\prime}{.}2179$ &  9.215 & 10.954 \\
id5901erq & 2016-07-25T16:58:19.5485 & $18^\textrm{h}53^\textrm{m}53\overset{\textrm{s}}{.}00295$ & $-20^\circ46^\prime32\overset{\prime\prime}{.}2817$ &  7.198 &  6.205 \\
id5901etq & 2016-07-25T17:07:07.5328 & $18^\textrm{h}53^\textrm{m}52\overset{\textrm{s}}{.}96820$ & $-20^\circ46^\prime32\overset{\prime\prime}{.}3433$ & 12.890 & 10.688 \\
id5901evq & 2016-07-25T17:15:55.5491 & $18^\textrm{h}53^\textrm{m}52\overset{\textrm{s}}{.}93097$ & $-20^\circ46^\prime32\overset{\prime\prime}{.}3865$ &  7.764 &  6.118 \\
id5901exq & 2016-07-25T17:24:43.5490 & $18^\textrm{h}53^\textrm{m}52\overset{\textrm{s}}{.}89789$ & $-20^\circ46^\prime32\overset{\prime\prime}{.}4503$ & 11.000 & 11.074 \\
id5902a9q & 2016-10-21T05:07:24.7290 & $18^\textrm{h}51^\textrm{m}48\overset{\textrm{s}}{.}31696$ & $-20^\circ53^\prime07\overset{\prime\prime}{.}6653$ &  9.090 &  9.960 \\
id5902adq & 2016-10-21T05:25:00.7297 & $18^\textrm{h}51^\textrm{m}48\overset{\textrm{s}}{.}33435$ & $-20^\circ53^\prime07\overset{\prime\prime}{.}5531$ & 10.424 &  8.911 \\
id5902afq & 2016-10-21T05:33:48.7295 & $18^\textrm{h}51^\textrm{m}48\overset{\textrm{s}}{.}34571$ & $-20^\circ53^\prime07\overset{\prime\prime}{.}4620$ &  8.163 &  8.154 \\
id5953gxq & 2017-05-01T21:02:43.4730 & $19^\textrm{h}05^\textrm{m}08\overset{\textrm{s}}{.}03529$ & $-20^\circ33^\prime13\overset{\prime\prime}{.}4643$ &  5.497 &  5.760 \\
id5953gyq & 2017-05-01T21:11:31.4574 & $19^\textrm{h}05^\textrm{m}08\overset{\textrm{s}}{.}02054$ & $-20^\circ33^\prime13\overset{\prime\prime}{.}4987$ &  6.192 &  5.165 \\
id5953h0q & 2017-05-01T21:20:19.4738 & $19^\textrm{h}05^\textrm{m}08\overset{\textrm{s}}{.}00391$ & $-20^\circ33^\prime13\overset{\prime\prime}{.}5663$ &  6.838 &  5.995 \\
id5953h2q & 2017-05-01T21:29:07.4736 & $19^\textrm{h}05^\textrm{m}07\overset{\textrm{s}}{.}98675$ & $-20^\circ33^\prime13\overset{\prime\prime}{.}6156$ &  7.427 &  7.120 \\
id5953h4q & 2017-05-01T21:37:55.4744 & $19^\textrm{h}05^\textrm{m}07\overset{\textrm{s}}{.}96987$ & $-20^\circ33^\prime13\overset{\prime\prime}{.}6559$ &  9.199 &  9.665 \\
id5904wiq & 2017-05-25T17:27:55.5436 & $19^\textrm{h}04^\textrm{m}06\overset{\textrm{s}}{.}07475$ & $-20^\circ34^\prime02\overset{\prime\prime}{.}9420$ &  6.379 &  5.984 \\
id5904wkq & 2017-05-25T17:36:43.5280 & $19^\textrm{h}04^\textrm{m}06\overset{\textrm{s}}{.}04412$ & $-20^\circ34^\prime02\overset{\prime\prime}{.}9634$ &  6.826 &  5.843 \\
id5904wmq & 2017-05-25T17:45:31.5434 & $19^\textrm{h}04^\textrm{m}06\overset{\textrm{s}}{.}01468$ & $-20^\circ34^\prime02\overset{\prime\prime}{.}9165$ &  6.391 &  7.571 \\
id8i01skq & 2017-06-25T07:37:23.6242 & $19^\textrm{h}01^\textrm{m}55\overset{\textrm{s}}{.}18310$ & $-20^\circ36^\prime53\overset{\prime\prime}{.}1436$ &  6.396 &  6.501 \\
id8i01slq & 2017-06-25T07:46:11.6241 & $19^\textrm{h}01^\textrm{m}55\overset{\textrm{s}}{.}14905$ & $-20^\circ36^\prime53\overset{\prime\prime}{.}2328$ &  6.156 &  6.246 \\
id8i01snq & 2017-06-25T07:54:59.6240 & $19^\textrm{h}01^\textrm{m}55\overset{\textrm{s}}{.}11119$ & $-20^\circ36^\prime53\overset{\prime\prime}{.}3237$ &  8.370 &  7.069 \\
id8i01spq & 2017-06-25T08:03:47.6231 & $19^\textrm{h}01^\textrm{m}55\overset{\textrm{s}}{.}07326$ & $-20^\circ36^\prime53\overset{\prime\prime}{.}4114$ &  9.158 &  7.480 \\
id8i01srq & 2017-06-25T08:12:35.6074 & $19^\textrm{h}01^\textrm{m}55\overset{\textrm{s}}{.}03946$ & $-20^\circ36^\prime53\overset{\prime\prime}{.}4787$ &  5.086 &  6.422 \\
id8i02stq & 2017-06-25T09:12:44.6149 & $19^\textrm{h}01^\textrm{m}54\overset{\textrm{s}}{.}86318$ & $-20^\circ36^\prime53\overset{\prime\prime}{.}5963$ &  9.097 &  7.171 \\
id8i02suq & 2017-06-25T09:21:32.6148 & $19^\textrm{h}01^\textrm{m}54\overset{\textrm{s}}{.}82820$ & $-20^\circ36^\prime53\overset{\prime\prime}{.}6883$ &  7.471 &  5.608 \\
id8i02swq & 2017-06-25T09:30:20.6156 & $19^\textrm{h}01^\textrm{m}54\overset{\textrm{s}}{.}79110$ & $-20^\circ36^\prime53\overset{\prime\prime}{.}7914$ &  4.644 &  4.964 \\
id8i02syq & 2017-06-25T09:39:08.6155 & $19^\textrm{h}01^\textrm{m}54\overset{\textrm{s}}{.}75385$ & $-20^\circ36^\prime53\overset{\prime\prime}{.}8649$ &  6.545 &  5.584 \\
id8i02t0q & 2017-06-25T09:47:56.6163 & $19^\textrm{h}01^\textrm{m}54\overset{\textrm{s}}{.}71922$ & $-20^\circ36^\prime53\overset{\prime\prime}{.}9495$ &  5.128 &  4.908 \\
id8i03t4q & 2017-06-25T10:48:05.6073 & $19^\textrm{h}01^\textrm{m}54\overset{\textrm{s}}{.}54329$ & $-20^\circ36^\prime54\overset{\prime\prime}{.}0349$ &  6.562 &  7.028 \\
id8i03t7q & 2017-06-25T10:56:53.6081 & $19^\textrm{h}01^\textrm{m}54\overset{\textrm{s}}{.}50911$ & $-20^\circ36^\prime54\overset{\prime\prime}{.}1633$ &  5.107 &  5.347 \\
id8i03t9q & 2017-06-25T11:05:41.6080 & $19^\textrm{h}01^\textrm{m}54\overset{\textrm{s}}{.}47159$ & $-20^\circ36^\prime54\overset{\prime\prime}{.}2478$ &  5.968 &  7.687 \\
id8i03tbq & 2017-06-25T11:14:29.6070 & $19^\textrm{h}01^\textrm{m}54\overset{\textrm{s}}{.}43409$ & $-20^\circ36^\prime54\overset{\prime\prime}{.}3285$ &  5.276 &  5.845 \\
id8i04tfq & 2017-06-25T12:23:27.6227 & $19^\textrm{h}01^\textrm{m}54\overset{\textrm{s}}{.}22363$ & $-20^\circ36^\prime54\overset{\prime\prime}{.}5187$ &  5.490 &  6.343 \\
id8i04tgq & 2017-06-25T12:32:15.6071 & $19^\textrm{h}01^\textrm{m}54\overset{\textrm{s}}{.}18806$ & $-20^\circ36^\prime54\overset{\prime\prime}{.}6009$ &  8.028 &  7.617 \\
id8i04tiq & 2017-06-25T12:41:03.6234 & $19^\textrm{h}01^\textrm{m}54\overset{\textrm{s}}{.}15199$ & $-20^\circ36^\prime54\overset{\prime\prime}{.}7131$ &  5.463 &  6.314 \\
id8i04tkq & 2017-06-25T12:49:51.6242 & $19^\textrm{h}01^\textrm{m}54\overset{\textrm{s}}{.}11435$ & $-20^\circ36^\prime54\overset{\prime\prime}{.}7939$ &  6.269 &  6.610 \\
id8i04tmq & 2017-06-25T12:58:39.6240 & $19^\textrm{h}01^\textrm{m}54\overset{\textrm{s}}{.}07993$ & $-20^\circ36^\prime54\overset{\prime\prime}{.}8631$ &  5.614 &  5.066 \\
id8i05txq & 2017-06-25T15:34:10.6003 & $19^\textrm{h}01^\textrm{m}53\overset{\textrm{s}}{.}58450$ & $-20^\circ36^\prime55\overset{\prime\prime}{.}4388$ &  5.972 &  6.391 \\
id8i05tyq & 2017-06-25T15:42:58.5993 & $19^\textrm{h}01^\textrm{m}53\overset{\textrm{s}}{.}54976$ & $-20^\circ36^\prime55\overset{\prime\prime}{.}5321$ &  6.780 &  5.843 \\
id8i05u0q & 2017-06-25T15:51:46.6156 & $19^\textrm{h}01^\textrm{m}53\overset{\textrm{s}}{.}51243$ & $-20^\circ36^\prime55\overset{\prime\prime}{.}6468$ &  5.993 &  6.287 \\
id8i05u2q & 2017-06-25T16:00:34.5991 & $19^\textrm{h}01^\textrm{m}53\overset{\textrm{s}}{.}47352$ & $-20^\circ36^\prime55\overset{\prime\prime}{.}7304$ &  8.008 &  7.278 \\
id8i05u4q & 2017-06-25T16:09:22.5999 & $19^\textrm{h}01^\textrm{m}53\overset{\textrm{s}}{.}43916$ & $-20^\circ36^\prime55\overset{\prime\prime}{.}7992$ &  7.822 &  8.149 \\
id8i06u6q & 2017-06-25T17:09:30.6155 & $19^\textrm{h}01^\textrm{m}53\overset{\textrm{s}}{.}26390$ & $-20^\circ36^\prime55\overset{\prime\prime}{.}9206$ &  8.896 &  9.491 \\
id8i06u7q & 2017-06-25T17:18:18.5998 & $19^\textrm{h}01^\textrm{m}53\overset{\textrm{s}}{.}23041$ & $-20^\circ36^\prime55\overset{\prime\prime}{.}9925$ &  7.080 &  6.282 \\
id8i06u9q & 2017-06-25T17:27:06.5989 & $19^\textrm{h}01^\textrm{m}53\overset{\textrm{s}}{.}19192$ & $-20^\circ36^\prime56\overset{\prime\prime}{.}1086$ &  8.232 &  6.485 \\
id8i06ubq & 2017-06-25T17:35:54.6152 & $19^\textrm{h}01^\textrm{m}53\overset{\textrm{s}}{.}15377$ & $-20^\circ36^\prime56\overset{\prime\prime}{.}2064$ &  8.378 &  7.502 \\
id8i06udq & 2017-06-25T17:44:42.5987 & $19^\textrm{h}01^\textrm{m}53\overset{\textrm{s}}{.}11963$ & $-20^\circ36^\prime56\overset{\prime\prime}{.}2669$ &  8.389 &  7.610 \\
id8i07crq & 2017-06-26T07:27:41.6079 & $19^\textrm{h}01^\textrm{m}50\overset{\textrm{s}}{.}38003$ & $-20^\circ37^\prime00\overset{\prime\prime}{.}1011$ &  6.654 &  7.075 \\
id8i07ctq & 2017-06-26T07:36:29.6078 & $19^\textrm{h}01^\textrm{m}50\overset{\textrm{s}}{.}34397$ & $-20^\circ37^\prime00\overset{\prime\prime}{.}1993$ &  8.378 &  7.315 \\
id8i07cwq & 2017-06-26T07:45:17.6068 & $19^\textrm{h}01^\textrm{m}50\overset{\textrm{s}}{.}30856$ & $-20^\circ37^\prime00\overset{\prime\prime}{.}3029$ &  5.666 &  6.112 \\
id8i07cyq & 2017-06-26T07:54:05.6068 & $19^\textrm{h}01^\textrm{m}50\overset{\textrm{s}}{.}27096$ & $-20^\circ37^\prime00\overset{\prime\prime}{.}4163$ &  5.126 &  6.084 \\
id8i07d1q & 2017-06-26T08:02:53.6066 & $19^\textrm{h}01^\textrm{m}50\overset{\textrm{s}}{.}23531$ & $-20^\circ37^\prime00\overset{\prime\prime}{.}4803$ &  5.526 &  5.040 \\
id8i08daq & 2017-06-26T09:03:03.6233 & $19^\textrm{h}01^\textrm{m}50\overset{\textrm{s}}{.}05904$ & $-20^\circ37^\prime00\overset{\prime\prime}{.}5910$ & 10.279 &  6.816 \\
id8i08dbq & 2017-06-26T09:11:51.6232 & $19^\textrm{h}01^\textrm{m}50\overset{\textrm{s}}{.}02460$ & $-20^\circ37^\prime00\overset{\prime\prime}{.}6563$ &  6.995 &  6.221 \\
id8i08ddq & 2017-06-26T09:20:39.6240 & $19^\textrm{h}01^\textrm{m}49\overset{\textrm{s}}{.}98730$ & $-20^\circ37^\prime00\overset{\prime\prime}{.}7742$ &  5.442 &  6.276 \\
id8i08dfq & 2017-06-26T09:29:27.6238 & $19^\textrm{h}01^\textrm{m}49\overset{\textrm{s}}{.}94929$ & $-20^\circ37^\prime00\overset{\prime\prime}{.}8552$ &  8.475 &  8.027 \\
id8i08diq & 2017-06-26T09:38:15.6229 & $19^\textrm{h}01^\textrm{m}49\overset{\textrm{s}}{.}91336$ & $-20^\circ37^\prime00\overset{\prime\prime}{.}9372$ &  5.958 &  5.866 \\
id8i09dkq & 2017-06-26T10:38:25.6075 & $19^\textrm{h}01^\textrm{m}49\overset{\textrm{s}}{.}73909$ & $-20^\circ37^\prime01\overset{\prime\prime}{.}0586$ &  5.690 &  6.374 \\
id8i09dlq & 2017-06-26T10:47:13.6075 & $19^\textrm{h}01^\textrm{m}49\overset{\textrm{s}}{.}70415$ & $-20^\circ37^\prime01\overset{\prime\prime}{.}1403$ &  5.856 &  7.255 \\
id8i09dnq & 2017-06-26T10:56:01.6083 & $19^\textrm{h}01^\textrm{m}49\overset{\textrm{s}}{.}66535$ & $-20^\circ37^\prime01\overset{\prime\prime}{.}2596$ &  6.223 &  5.913 \\
id8i09dpq & 2017-06-26T11:04:49.6081 & $19^\textrm{h}01^\textrm{m}49\overset{\textrm{s}}{.}62903$ & $-20^\circ37^\prime01\overset{\prime\prime}{.}3395$ &  6.642 &  8.355 \\
id8i10dtq & 2017-06-26T12:13:46.6155 & $19^\textrm{h}01^\textrm{m}49\overset{\textrm{s}}{.}41640$ & $-20^\circ37^\prime01\overset{\prime\prime}{.}5098$ &  6.808 &  7.171 \\
id8i10duq & 2017-06-26T12:22:34.5990 & $19^\textrm{h}01^\textrm{m}49\overset{\textrm{s}}{.}38307$ & $-20^\circ37^\prime01\overset{\prime\prime}{.}6174$ &  5.280 &  6.514 \\
id8i10dwq & 2017-06-26T12:31:22.5989 & $19^\textrm{h}01^\textrm{m}49\overset{\textrm{s}}{.}34460$ & $-20^\circ37^\prime01\overset{\prime\prime}{.}7232$ &  6.411 &  6.077 \\
id8i10dyq & 2017-06-26T12:40:10.5997 & $19^\textrm{h}01^\textrm{m}49\overset{\textrm{s}}{.}30648$ & $-20^\circ37^\prime01\overset{\prime\prime}{.}8035$ &  4.767 &  5.779 \\
id8i10e0q & 2017-06-26T12:48:58.5996 & $19^\textrm{h}01^\textrm{m}49\overset{\textrm{s}}{.}27112$ & $-20^\circ37^\prime01\overset{\prime\prime}{.}8889$ &  5.476 &  5.328 \\
id8i11eiq & 2017-06-26T17:00:38.6153 & $19^\textrm{h}01^\textrm{m}48\overset{\textrm{s}}{.}45153$ & $-20^\circ37^\prime02\overset{\prime\prime}{.}9441$ &  8.482 &  8.030 \\
id8i11ejq & 2017-06-26T17:09:26.5988 & $19^\textrm{h}01^\textrm{m}48\overset{\textrm{s}}{.}41519$ & $-20^\circ37^\prime03\overset{\prime\prime}{.}0318$ &  8.076 &  7.328 \\
id8i11elq & 2017-06-26T17:18:14.5987 & $19^\textrm{h}01^\textrm{m}48\overset{\textrm{s}}{.}37703$ & $-20^\circ37^\prime03\overset{\prime\prime}{.}1593$ &  6.787 &  9.734 \\
id8i11eoq & 2017-06-26T17:27:02.5995 & $19^\textrm{h}01^\textrm{m}48\overset{\textrm{s}}{.}33962$ & $-20^\circ37^\prime03\overset{\prime\prime}{.}2174$ &  7.868 & 10.419 \\
id8i11eqq & 2017-06-26T17:35:50.5993 & $19^\textrm{h}01^\textrm{m}48\overset{\textrm{s}}{.}30633$ & $-20^\circ37^\prime03\overset{\prime\prime}{.}2853$ &  5.290 &  4.749 \\
id8i12esq & 2017-06-26T18:40:22.5998 & $19^\textrm{h}01^\textrm{m}48\overset{\textrm{s}}{.}11203$ & $-20^\circ37^\prime03\overset{\prime\prime}{.}4625$ &  6.732 &  5.483 \\
id8i12etq & 2017-06-26T18:49:10.5997 & $19^\textrm{h}01^\textrm{m}48\overset{\textrm{s}}{.}07483$ & $-20^\circ37^\prime03\overset{\prime\prime}{.}5526$ &  7.972 &  8.115 \\
id8i12evq & 2017-06-26T18:57:58.6152 & $19^\textrm{h}01^\textrm{m}48\overset{\textrm{s}}{.}03824$ & $-20^\circ37^\prime03\overset{\prime\prime}{.}6610$ &  8.285 &  6.559 \\
id8i12exq & 2017-06-26T19:06:46.5987 & $19^\textrm{h}01^\textrm{m}48\overset{\textrm{s}}{.}00234$ & $-20^\circ37^\prime03\overset{\prime\prime}{.}7452$ &  6.643 &  8.809 \\
id8i13lvq & 2017-06-28T05:32:56.6157 & $19^\textrm{h}01^\textrm{m}41\overset{\textrm{s}}{.}04678$ & $-20^\circ37^\prime13\overset{\prime\prime}{.}8017$ &  6.423 &  5.150 \\
id8i13lwq & 2017-06-28T05:41:44.6156 & $19^\textrm{h}01^\textrm{m}41\overset{\textrm{s}}{.}01136$ & $-20^\circ37^\prime13\overset{\prime\prime}{.}9082$ &  5.077 &  4.970 \\
id8i13lyq & 2017-06-28T05:50:32.6155 & $19^\textrm{h}01^\textrm{m}40\overset{\textrm{s}}{.}97517$ & $-20^\circ37^\prime14\overset{\prime\prime}{.}0037$ &  4.861 &  4.601 \\
id8i13m0q & 2017-06-28T05:59:20.6163 & $19^\textrm{h}01^\textrm{m}40\overset{\textrm{s}}{.}93805$ & $-20^\circ37^\prime14\overset{\prime\prime}{.}1005$ &  6.567 &  6.656 \\
id8i13m3q & 2017-06-28T06:08:08.6161 & $19^\textrm{h}01^\textrm{m}40\overset{\textrm{s}}{.}90205$ & $-20^\circ37^\prime14\overset{\prime\prime}{.}1767$ &  6.510 &  5.490 \\
id8i14m5q & 2017-06-28T07:08:17.6081 & $19^\textrm{h}01^\textrm{m}40\overset{\textrm{s}}{.}72466$ & $-20^\circ37^\prime14\overset{\prime\prime}{.}2761$ &  4.571 &  4.725 \\
id8i14m6q & 2017-06-28T07:17:05.6071 & $19^\textrm{h}01^\textrm{m}40\overset{\textrm{s}}{.}68925$ & $-20^\circ37^\prime14\overset{\prime\prime}{.}3838$ &  5.149 &  5.421 \\
id8i14m8q & 2017-06-28T07:25:53.6070 & $19^\textrm{h}01^\textrm{m}40\overset{\textrm{s}}{.}65300$ & $-20^\circ37^\prime14\overset{\prime\prime}{.}4841$ &  6.629 &  6.663 \\
id8i14maq & 2017-06-28T07:34:41.6069 & $19^\textrm{h}01^\textrm{m}40\overset{\textrm{s}}{.}61368$ & $-20^\circ37^\prime14\overset{\prime\prime}{.}6007$ &  5.686 &  6.738 \\
id8i14mcq & 2017-06-28T07:43:29.6077 & $19^\textrm{h}01^\textrm{m}40\overset{\textrm{s}}{.}58000$ & $-20^\circ37^\prime14\overset{\prime\prime}{.}6548$ &  9.290 &  7.692 \\
id8i15meq & 2017-06-28T08:43:37.6078 & $19^\textrm{h}01^\textrm{m}40\overset{\textrm{s}}{.}40122$ & $-20^\circ37^\prime14\overset{\prime\prime}{.}7501$ &  6.204 &  5.847 \\
id8i15mfq & 2017-06-28T08:52:25.6076 & $19^\textrm{h}01^\textrm{m}40\overset{\textrm{s}}{.}36750$ & $-20^\circ37^\prime14\overset{\prime\prime}{.}8581$ &  6.689 &  5.987 \\
id8i15mhq & 2017-06-28T09:01:13.6067 & $19^\textrm{h}01^\textrm{m}40\overset{\textrm{s}}{.}33020$ & $-20^\circ37^\prime14\overset{\prime\prime}{.}9650$ &  4.155 &  4.586 \\
id8i15mjq & 2017-06-28T09:10:01.6075 & $19^\textrm{h}01^\textrm{m}40\overset{\textrm{s}}{.}29244$ & $-20^\circ37^\prime15\overset{\prime\prime}{.}0795$ &  5.379 &  6.786 \\
id8i15mlq & 2017-06-28T09:18:49.6082 & $19^\textrm{h}01^\textrm{m}40\overset{\textrm{s}}{.}25699$ & $-20^\circ37^\prime15\overset{\prime\prime}{.}1355$ &  5.312 &  5.183 \\
id8i16mnq & 2017-06-28T10:18:59.6232 & $19^\textrm{h}01^\textrm{m}40\overset{\textrm{s}}{.}07761$ & $-20^\circ37^\prime15\overset{\prime\prime}{.}2312$ &  8.361 &  9.052 \\
id8i16moq & 2017-06-28T10:27:47.6230 & $19^\textrm{h}01^\textrm{m}40\overset{\textrm{s}}{.}04398$ & $-20^\circ37^\prime15\overset{\prime\prime}{.}3414$ &  5.624 &  5.232 \\
id8i16mqq & 2017-06-28T10:36:35.6238 & $19^\textrm{h}01^\textrm{m}40\overset{\textrm{s}}{.}00812$ & $-20^\circ37^\prime15\overset{\prime\prime}{.}4442$ &  5.787 &  5.530 \\
id8i16msq & 2017-06-28T10:45:23.6237 & $19^\textrm{h}01^\textrm{m}39\overset{\textrm{s}}{.}96933$ & $-20^\circ37^\prime15\overset{\prime\prime}{.}5235$ &  7.036 &  6.210 \\
id8i16muq & 2017-06-28T10:54:11.6236 & $19^\textrm{h}01^\textrm{m}39\overset{\textrm{s}}{.}93349$ & $-20^\circ37^\prime15\overset{\prime\prime}{.}6005$ &  6.019 &  6.975 \\
id8i17obq & 2017-06-28T15:05:02.6151 & $19^\textrm{h}01^\textrm{m}39\overset{\textrm{s}}{.}10997$ & $-20^\circ37^\prime16\overset{\prime\prime}{.}6579$ &  7.323 &  5.389 \\
id8i17ocq & 2017-06-28T15:13:50.6003 & $19^\textrm{h}01^\textrm{m}39\overset{\textrm{s}}{.}07618$ & $-20^\circ37^\prime16\overset{\prime\prime}{.}7759$ &  4.671 &  4.676 \\
id8i17oeq & 2017-06-28T15:22:38.6002 & $19^\textrm{h}01^\textrm{m}39\overset{\textrm{s}}{.}03924$ & $-20^\circ37^\prime16\overset{\prime\prime}{.}8792$ &  6.235 &  6.510 \\
id8i17ogq & 2017-06-28T15:31:26.6157 & $19^\textrm{h}01^\textrm{m}39\overset{\textrm{s}}{.}00035$ & $-20^\circ37^\prime16\overset{\prime\prime}{.}9833$ &  6.294 &  7.602 \\
id8i17oiq & 2017-06-28T15:40:14.5992 & $19^\textrm{h}01^\textrm{m}38\overset{\textrm{s}}{.}96504$ & $-20^\circ37^\prime17\overset{\prime\prime}{.}0334$ &  5.759 &  4.966 \\
id8i18olq & 2017-06-28T16:49:11.6238 & $19^\textrm{h}01^\textrm{m}38\overset{\textrm{s}}{.}75129$ & $-20^\circ37^\prime17\overset{\prime\prime}{.}2404$ &  6.936 &  5.735 \\
id8i18onq & 2017-06-28T16:57:59.6237 & $19^\textrm{h}01^\textrm{m}38\overset{\textrm{s}}{.}71440$ & $-20^\circ37^\prime17\overset{\prime\prime}{.}3650$ &  6.058 &  5.273 \\
id8i18opq & 2017-06-28T17:06:47.6228 & $19^\textrm{h}01^\textrm{m}38\overset{\textrm{s}}{.}67779$ & $-20^\circ37^\prime17\overset{\prime\prime}{.}4586$ &  5.509 &  5.146 \\
id8i18orq & 2017-06-28T17:15:35.6071 & $19^\textrm{h}01^\textrm{m}38\overset{\textrm{s}}{.}64165$ & $-20^\circ37^\prime17\overset{\prime\prime}{.}5070$ &  6.706 &  5.493 \\
id8i19geq & 2017-07-04T04:34:39.6076 & $19^\textrm{h}01^\textrm{m}11\overset{\textrm{s}}{.}82432$ & $-20^\circ37^\prime57\overset{\prime\prime}{.}6469$ &  6.321 &  8.507 \\
id8i19gfq & 2017-07-04T04:43:27.6239 & $19^\textrm{h}01^\textrm{m}11\overset{\textrm{s}}{.}79038$ & $-20^\circ37^\prime57\overset{\prime\prime}{.}7960$ &  7.592 &  8.341 \\
id8i19gpq & 2017-07-04T04:52:15.6074 & $19^\textrm{h}01^\textrm{m}11\overset{\textrm{s}}{.}75312$ & $-20^\circ37^\prime57\overset{\prime\prime}{.}9103$ &  6.640 &  7.705 \\
id8i19grq & 2017-07-04T05:01:03.6229 & $19^\textrm{h}01^\textrm{m}11\overset{\textrm{s}}{.}71403$ & $-20^\circ37^\prime57\overset{\prime\prime}{.}9920$ &  9.998 &  5.635 \\
id8i19gtq & 2017-07-04T05:09:51.6228 & $19^\textrm{h}01^\textrm{m}11\overset{\textrm{s}}{.}67763$ & $-20^\circ37^\prime58\overset{\prime\prime}{.}0126$ &  9.165 &  7.546 \\
id8i20h2q & 2017-07-04T06:10:00.6147 & $19^\textrm{h}01^\textrm{m}11\overset{\textrm{s}}{.}49739$ & $-20^\circ37^\prime58\overset{\prime\prime}{.}1574$ &  7.833 &  7.575 \\
id8i20h3q & 2017-07-04T06:18:48.6155 & $19^\textrm{h}01^\textrm{m}11\overset{\textrm{s}}{.}46445$ & $-20^\circ37^\prime58\overset{\prime\prime}{.}2853$ &  6.418 &  5.336 \\
id8i20h5q & 2017-07-04T06:27:36.6154 & $19^\textrm{h}01^\textrm{m}11\overset{\textrm{s}}{.}42716$ & $-20^\circ37^\prime58\overset{\prime\prime}{.}4153$ &  5.600 &  5.271 \\
id8i20h7q & 2017-07-04T06:36:24.6153 & $19^\textrm{h}01^\textrm{m}11\overset{\textrm{s}}{.}38824$ & $-20^\circ37^\prime58\overset{\prime\prime}{.}4798$ &  6.678 &  7.588 \\
id8i20h9q & 2017-07-04T06:45:12.6161 & $19^\textrm{h}01^\textrm{m}11\overset{\textrm{s}}{.}35226$ & $-20^\circ37^\prime58\overset{\prime\prime}{.}5298$ &  7.448 &  7.369 \\
id8i21hbq & 2017-07-04T07:45:23.6237 & $19^\textrm{h}01^\textrm{m}11\overset{\textrm{s}}{.}17047$ & $-20^\circ37^\prime58\overset{\prime\prime}{.}6434$ &  8.945 & 11.126 \\
id8i21hcq & 2017-07-04T07:54:11.6236 & $19^\textrm{h}01^\textrm{m}11\overset{\textrm{s}}{.}13786$ & $-20^\circ37^\prime58\overset{\prime\prime}{.}7765$ &  6.935 &  8.339 \\
id8i21hgq & 2017-07-04T08:11:47.6243 & $19^\textrm{h}01^\textrm{m}11\overset{\textrm{s}}{.}06238$ & $-20^\circ37^\prime58\overset{\prime\prime}{.}9846$ & 10.580 & 10.329 \\
id8i21hiq & 2017-07-04T08:20:35.6078 & $19^\textrm{h}01^\textrm{m}11\overset{\textrm{s}}{.}02544$ & $-20^\circ37^\prime59\overset{\prime\prime}{.}0383$ &  6.797 &  6.626 \\
id8i22hyq & 2017-07-04T12:31:26.6001 & $19^\textrm{h}01^\textrm{m}10\overset{\textrm{s}}{.}19388$ & $-20^\circ38^\prime00\overset{\prime\prime}{.}1481$ &  5.143 &  6.784 \\
id8i22hzq & 2017-07-04T12:40:14.5992 & $19^\textrm{h}01^\textrm{m}10\overset{\textrm{s}}{.}16130$ & $-20^\circ38^\prime00\overset{\prime\prime}{.}2791$ &  8.660 &  8.245 \\
id8i22i1q & 2017-07-04T12:49:02.6155 & $19^\textrm{h}01^\textrm{m}10\overset{\textrm{s}}{.}12266$ & $-20^\circ38^\prime00\overset{\prime\prime}{.}3949$ &  5.418 &  4.429 \\
id8i22i3q & 2017-07-04T12:57:50.5998 & $19^\textrm{h}01^\textrm{m}10\overset{\textrm{s}}{.}08322$ & $-20^\circ38^\prime00\overset{\prime\prime}{.}4893$ &  7.182 &  6.729 \\
id8i22i5q & 2017-07-04T13:06:38.5997 & $19^\textrm{h}01^\textrm{m}10\overset{\textrm{s}}{.}04743$ & $-20^\circ38^\prime00\overset{\prime\prime}{.}5278$ &  3.997 &  4.890 \\
id8i23i7q & 2017-07-04T14:06:48.6155 & $19^\textrm{h}01^\textrm{m}09\overset{\textrm{s}}{.}86719$ & $-20^\circ38^\prime00\overset{\prime\prime}{.}6361$ &  4.596 &  6.556 \\
id8i23i8q & 2017-07-04T14:15:36.6163 & $19^\textrm{h}01^\textrm{m}09\overset{\textrm{s}}{.}83328$ & $-20^\circ38^\prime00\overset{\prime\prime}{.}7834$ &  5.196 &  5.465 \\
id8i23iaq & 2017-07-04T14:24:24.6162 & $19^\textrm{h}01^\textrm{m}09\overset{\textrm{s}}{.}79710$ & $-20^\circ38^\prime00\overset{\prime\prime}{.}9007$ &  6.766 &  5.229 \\
id8i23icq & 2017-07-04T14:33:12.6152 & $19^\textrm{h}01^\textrm{m}09\overset{\textrm{s}}{.}75866$ & $-20^\circ38^\prime00\overset{\prime\prime}{.}9762$ &  8.809 &  8.754 \\
id8i23ieq & 2017-07-04T14:42:00.6151 & $19^\textrm{h}01^\textrm{m}09\overset{\textrm{s}}{.}72136$ & $-20^\circ38^\prime01\overset{\prime\prime}{.}0241$ &  7.927 &  8.082 \\
id8i24igq & 2017-07-04T15:42:09.6071 & $19^\textrm{h}01^\textrm{m}09\overset{\textrm{s}}{.}54125$ & $-20^\circ38^\prime01\overset{\prime\prime}{.}1456$ &  6.294 &  6.857 \\
id8i24ihq & 2017-07-04T15:50:57.6069 & $19^\textrm{h}01^\textrm{m}09\overset{\textrm{s}}{.}50862$ & $-20^\circ38^\prime01\overset{\prime\prime}{.}2735$ &  6.219 &  5.827 \\
id8i24ijq & 2017-07-04T15:59:45.6077 & $19^\textrm{h}01^\textrm{m}09\overset{\textrm{s}}{.}47080$ & $-20^\circ38^\prime01\overset{\prime\prime}{.}3997$ &  7.332 &  5.674 \\
id8i24ilq & 2017-07-04T16:08:33.6076 & $19^\textrm{h}01^\textrm{m}09\overset{\textrm{s}}{.}43148$ & $-20^\circ38^\prime01\overset{\prime\prime}{.}4710$ &  6.157 &  8.619 \\
id8i24inq & 2017-07-04T16:17:21.6075 & $19^\textrm{h}01^\textrm{m}09\overset{\textrm{s}}{.}39554$ & $-20^\circ38^\prime01\overset{\prime\prime}{.}5173$ & 10.190 &  7.668 \\
id5906kaq & 2017-08-19T14:52:39.1725 & $18^\textrm{h}57^\textrm{m}47\overset{\textrm{s}}{.}65237$ & $-20^\circ43^\prime53\overset{\prime\prime}{.}8978$ &  7.664 &  8.171 \\
id5906kbq & 2017-08-19T15:01:27.1724 & $18^\textrm{h}57^\textrm{m}47\overset{\textrm{s}}{.}62571$ & $-20^\circ43^\prime54\overset{\prime\prime}{.}0078$ &  6.409 &  5.645 \\
id5906kdq & 2017-08-19T15:10:15.1731 & $18^\textrm{h}57^\textrm{m}47\overset{\textrm{s}}{.}59611$ & $-20^\circ43^\prime54\overset{\prime\prime}{.}1052$ &  8.899 &  8.690 \\
id5906khq & 2017-08-19T15:27:51.1721 & $18^\textrm{h}57^\textrm{m}47\overset{\textrm{s}}{.}54598$ & $-20^\circ43^\prime54\overset{\prime\prime}{.}2293$ &  7.351 &  7.481 \\
idoy07ytq & 2017-10-27T13:27:02.7694 & $18^\textrm{h}57^\textrm{m}29\overset{\textrm{s}}{.}93476$ & $-20^\circ47^\prime42\overset{\prime\prime}{.}9316$ & 11.725 & 10.181 \\
idoy07yuq & 2017-10-27T13:35:50.7693 & $18^\textrm{h}57^\textrm{m}29\overset{\textrm{s}}{.}94745$ & $-20^\circ47^\prime42\overset{\prime\prime}{.}9855$ &  7.217 &  7.990 \\
idoy07ywq & 2017-10-27T13:44:38.7692 & $18^\textrm{h}57^\textrm{m}29\overset{\textrm{s}}{.}95599$ & $-20^\circ47^\prime43\overset{\prime\prime}{.}0015$ & 11.805 &  9.046 \\
idoy07yyq & 2017-10-27T13:53:26.7699 & $18^\textrm{h}57^\textrm{m}29\overset{\textrm{s}}{.}96779$ & $-20^\circ47^\prime42\overset{\prime\prime}{.}9520$ &  7.191 &  7.225 \\
idoy07z0q & 2017-10-27T14:02:14.7698 & $18^\textrm{h}57^\textrm{m}29\overset{\textrm{s}}{.}98189$ & $-20^\circ47^\prime42\overset{\prime\prime}{.}8604$ &  8.738 &  9.698 \\

\enddata
\end{deluxetable*}

\end{document}